\documentclass{iopjournal}

\usepackage{amsmath,amssymb}
\usepackage{bm}

\newcommand{\Ntot}{N_{\text{tot}}}

\begin{document}

\articletype{Paper} 

\title{Yielding versus random organization: convex absorbing transitions in soft matter}

\author{Tristan Jocteur$^1$\orcid{0009-0002-6084-5788}, Kirsten Martens$^1$\orcid{0000-0003-2985-7385}, Eric Bertin$^1$\orcid{0000-0002-7354-6755} and Romain Mari$^{1,*}$\orcid{0000-0001-7877-416X}}

\affil{$^1$Univ.~Grenoble-Alpes, CNRS, LIPhy, 38000 Grenoble, France}

\affil{$^*$Author to whom any correspondence should be addressed.}

\email{romain.mari@univ-grenoble-alpes.fr}


\begin{abstract}
We explore the similarities and differences in the behavior of two different soft matter models, a generalized Random Organization Model (ROM) describing the stroboscopic dynamics of cyclically sheared suspensions, and an elastoplastic model describing the mesoscale dynamics of a yield-stress fluid under imposed stress.
Both systems can be described in the framework of absorbing phase transitions, which are nonequilibrium phase transitions between an absorbing (frozen) phase and an active phase.
In both models, a peculiar mechanism is at work: activity induces an internal noise which is transmitted over large distances by long-ranged mediated interactions, either hydrodynamic or elastic, which results in non-local creation of activity. 
This microscopic similarity is echoed in the critical behavior, as in both cases the transitions are convex, i.e., the order parameter (here the mean activity) is convex as a function of the control parameter (i.e., the exponent $\beta >1$). 
This is in stark contrast with usual absorbing phase transition, like (Conserved) Directed Percolation, which are concave ($\beta <1$).
Taking the power-law decay exponent $\alpha$ of long-range interactions as a control parameter, we compare the dependence of the critical properties
(activity mean value and fluctuations, avalanche statistics, low-wavenumber structure factor) on the decay exponent $\alpha$ in both models, finding a qualitatively similar scenario. 
A smooth crossover is observed as a function of $\alpha$ between a concave transition regime for short-range interactions, with diverging fluctuations and compact avalanches, and a convex transition regime, with vanishing fluctuations and non-compact avalanches, for longer-range interactions.
Although for a given range exponent $\alpha$, the values of critical exponents for both models differ, a good agreement between the models is found by parametrically plotting the different critical exponents as a function of the exponent $\beta$ of the mean activity.
In this parametric representation, the concave regime is consistent with the behavior of the Long-Range Directed Percolation class,
while the convex regime can be accounted for by a mean-field-type scenario with anomalous diffusion close to an absorbing boundary, inspired by the H\'ebraud-Lequeux model for the yielding transition.
\end{abstract}

\section{Introduction}

Absorbing phase transitions (APTs) form an important class of nonequilibrium phase transitions, whereby a system smoothly goes from a fluctuating active state to an absorbing state in which the dynamics is trapped \cite{hinrichsenNonequilibriumCriticalPhenomena2000,lubeckUniversalScalingBehavior2004}. Examples of absorbing phase transitions can be found in many different areas, ranging e.g.~from fracture propagation to epidemics~\cite{marroNonequilibriumPhaseTransitions1999,hinrichsenNonequilibriumPhaseTransitions2006,henkelAbsorbingPhaseTransitions2008}.
In a soft matter context, absorbing phase transitions are the relevant theoretical framework to describe
the reversible-irreversible transition (RIT) in oscillatory sheared suspensions of particles 
\cite{pineChaosThresholdIrreversibility2005,corteRandomOrganizationPeriodically2008,corteSelfOrganizedCriticalitySheared2009,menonUniversalityClassReversibleirreversible2009,tjhungHyperuniformDensityFluctuations2015,nessAbsorbingStateTransitionsGranular2020,gallianoTwoDimensionalCrystalsFar2023,bhowmikAbsorbingstateTransitionsParticulate2025,agrawalRheologyDynamicsDense2025}, vibrated granular media \cite{maireInterplayAbsorbingPhase2024,maireDynamicalStructuralProperties2025}, or the yielding transition in complex fluids under shear \cite{jocteurYieldingAbsorbingPhase2024a,nicolas_deformation_2018,coussotYieldStressFluid2014,bonn_yield_2017}.

APTs are usually classified into a few known universality classes \cite{lubeckUniversalScalingBehavior2004,henkelAbsorbingPhaseTransitions2008}.
In particular, systems with short-range interactions displaying an APT with infinitely many absorbing states unrelated by symmetry due to the presence of a conserved quantity, are considered to belong to the Conserved Directed Percolation (CDP) universality class~\cite{hinrichsenNonequilibriumCriticalPhenomena2000,mannaTwostateModelSelforganized1991,vespignani1998driving,rossi2000universality,le2015exact},
whose critical behavior has been well characterized.
In particular, the CDP universality class corresponds to a concave transition, in the sense that the order parameter (here the mean activity) displays a concave curve as a function of the control parameter (i.e., the critical exponent $\beta<1$).
A noticeable example, the depinning transition occurring in driven elastic manifolds in random media ~\cite{koplikInterfaceMovingRandom1985,nattermannDynamicsInterfaceDepinning1992,narayanThresholdCriticalDynamics1993,kardarNonequilibriumDynamicsInterfaces1998}
has been shown to belong to the CDP class \cite{alavaQuenchedNoiseOveractive2001,bonachelaAbsorbingStatesElastic2007,le2015exact}.

\begin{figure}[h]
    \centering
    \includegraphics[width=0.8\linewidth]{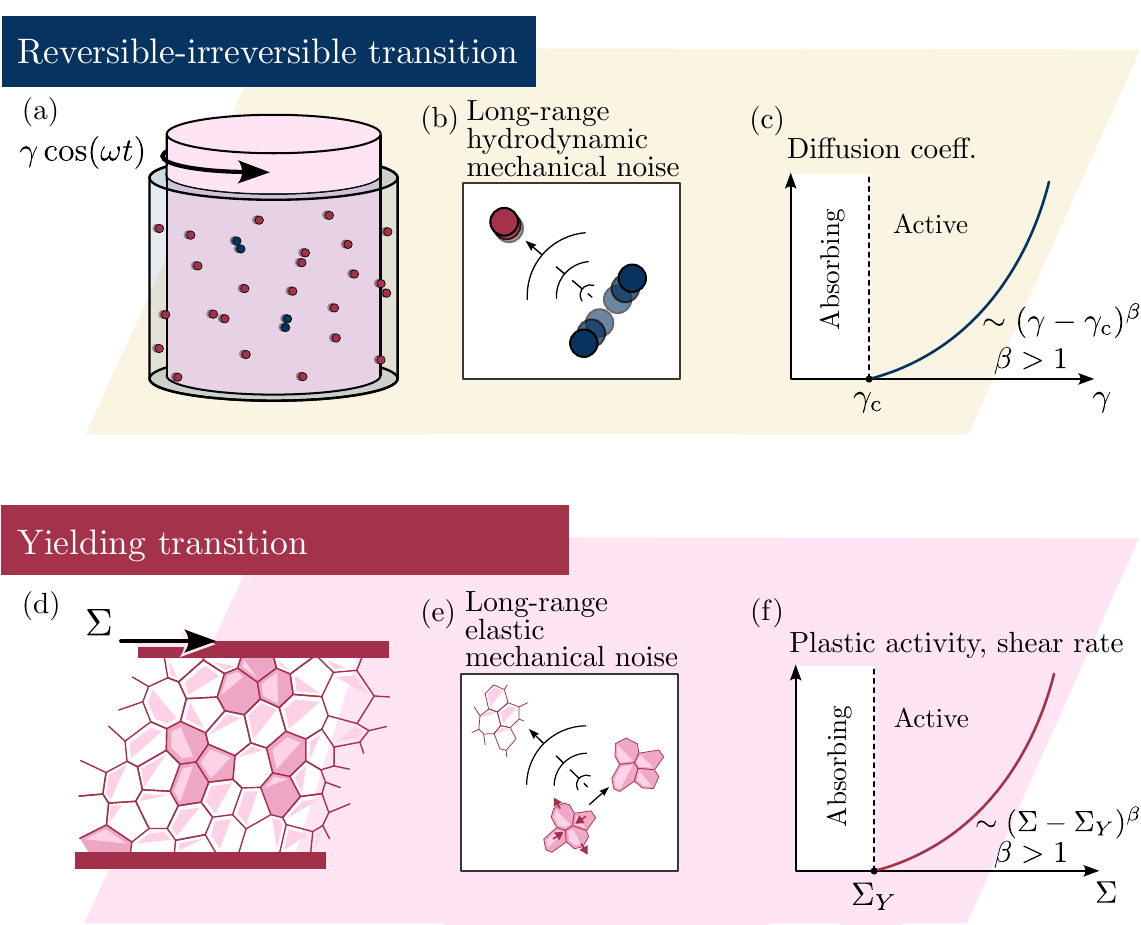}
    \caption{Similarities between the reversible-irreversible transition (RIT) of cyclically sheared suspensions and the yielding transition of yield stress fluids. Top: RIT. (a) A suspension of neutrally buoyant particles is cyclically sheared in a cylindrical Couette cell, and the particle positions are observed stroboscopically, once per cycle. If no interparticle contact occurs during the shear cycle, the particles follow reversible trajectories, and appear static in the stroboscopic sampling, i.e. the suspension is in an absorbing state. (b) If a contact occurs, there is an apparent diffusion in the stroboscopic dynamics for both the particles involved in a contact \emph{and} the surrounding ones, due to hydrodynamically mediated mechanical noise. (c) The absorbing phase transition when varying the shear amplitude $\gamma$, which separates an absorbing phase for $\gamma<\gamma_\mathrm{c}$ from an active phase for $\gamma>\gamma_\mathrm{c}$, is convex, $\beta>1$.
    Bottom: yielding transition. (d) A yield stress fluid (here a foam for illustrative purposes) is subjected to a shear stress $\Sigma$. Under stress, some parts of the sample deform elastically (lighter shade of pink) while localized parts turn plastic (darker shade). (e) When plasticity occurs, the stress relaxes in the plastic region, but it also propagates in the surrounding elastic medium, due to elastically mediated mechanical noise. (f) The absorbing phase transition when varying the applied stress $\Sigma$, which separates a static absorbing phase for $\Sigma < \Sigma_Y$ from a flowing active phase for $\Sigma > \Sigma_Y$, is convex.}
    \label{fig:resume}
\end{figure}

In a soft matter context, two paradigmatic transitions however might defy this classification, showing in particular a convex behavior ($\beta>1$) in presence of  long-range interactions.
The first one is a suspension of non-Brownian hard particles below jamming, subjected to cyclic shear, sketched in Fig.~\ref{fig:resume}(a).
When driven by an oscillatory shear of sufficiently small amplitude $\gamma$, the system eventually reaches a limit cycle after a finite number of deformation cycles, such that the microscopic configuration repeatedly follows the same trajectory.
As a result, when the system is observed stroboscopically, i.e. once per cycle, the dynamics appears fully arrested and the microscopic configuration remains effectively frozen.
This behavior, however, only persists below a critical driving amplitude $\gamma_c$ [Fig.~\ref{fig:resume}(c)].
Above this threshold, the stroboscopic dynamics remains diffusive even after an arbitrarily large number of cycles.

First identified in non-Brownian suspensions~\cite{pineChaosThresholdIrreversibility2005,corteRandomOrganizationPeriodically2008,corteSelfOrganizedCriticalitySheared2009}, this phenomenon, known as the Reversible-Irreversible Transition (RIT), has subsequently been reported experimentally in a broad variety of athermal complex fluids, including dry granular materials~\cite{royerPreciselyCyclicSand2015}, microemulsions~\cite{weijsEmergentHyperuniformityPeriodically2015}, foams~\cite{mukherjiStrengthMechanicalMemories2019}, soft glasses~\cite{fioccoOscillatoryAthermalQuasistatic2013}, or crumpled paper~\cite{shohatMemoryCoupledInstabilities2022}.
The phenomenon is also routinely reported in simulations or models of disordered athermal systems~\cite{milzConnectingRandomOrganization2013,leiHydrodynamicsRandomorganizingHyperuniform2019,nessAbsorbingStateTransitionsGranular2020,munganCyclicAnnealingIterated2019,munganNetworksHierarchiesHow2019,wilkenRandomClosePacking2021,liuFateShearoscillatedAmorphous2022,guoActiveDiffusingCrystals2025,huangBandingPolarizationDriven2025,gallianoGlassJammingTransitions2026,anandEmergentUniversalLongrange2026}.
Despite the microscopic diversity of systems displaying RIT, a common minimal phenomenology emerges.
In all known cases, irreversibility during a shear cycle originates from discrete and localized irreversible events, rather than from the exponential sensitivity characteristic of chaotic dynamical systems.
In unjammed suspensions, such irreversible events correspond to particle contacts occurring during the cycle [Fig.~\ref{fig:resume}(b)], which break the time-reversal symmetry otherwise guaranteed by Stokes hydrodynamics~\cite{corteRandomOrganizationPeriodically2008}.

The RIT is an APT, where limit cycles for $\gamma<\gamma_c$ are the absorbing states, and the diffusing dynamics for $\gamma>\gamma_c$ is the active phase~\cite{pineChaosThresholdIrreversibility2005,corteRandomOrganizationPeriodically2008,corteSelfOrganizedCriticalitySheared2009}.
Quantifying the activity $A$ using the stroboscopic diffusion coefficient of the suspended particles $D$, one finds for $\gamma>\gamma_c$ a critical behavior $A\equiv D\sim (\gamma-\gamma_c)^\beta$, Fig.~\ref{fig:resume}(c).
The RIT shares key properties with the CDP class: there is \emph{(i)} a conserved quantity (the density of particles) coupled to the activity (a higher density makes contacts more likely), and \emph{(ii)} an infinite number of absorbing states (all particle configurations that do not lead to any contact during the shear cycle). The first proposed simplified minimal models for the RIT indeed recovered the CDP critical behavior~\cite{corteRandomOrganizationPeriodically2008,menonUniversalityClassReversibleirreversible2009}. 
However, upon inclusion of long-range interactions induced by localized contact forces and mediated by the surrounding fluid, it was found that the transition shows a criticality different from CDP, with $\beta>1$~\cite{mariAbsorbingPhaseTransitions2022, jocteurRandom2025, jocteurAvalanches2026}.

The second paradigmatic example of a system exhibiting arrested dynamics and absorbing states is the yielding transition of yield stress fluids, sketched in Fig.~\ref{fig:resume}(d).
Yield stress fluids comprise a wide range of soft amorphous materials, such as emulsions, foams, and gels~\cite{coussotYieldStressFluid2014,bonn_yield_2017}.
When subjected to an external mechanical stress $\Sigma$, they respond as visco-elastic solids for stresses below a critical yield stress $\Sigma_Y$, while for $\Sigma>\Sigma_Y$ they flow as visco-elastic fluids.
In steady state, their rheological behavior is commonly described by the Herschel--Bulkley relation
$\Sigma = \Sigma_Y + k \dot\gamma^n$~\cite{herschelKonsistenzmessungenGummiBenzollosungen1926}, where $\dot\gamma$ denotes the shear rate, $k>0$, and $0<n\leq1$
~\cite{princenRheologyFoamsHighly1989,masonYieldingFlowMonodisperse1996,robertsNewMeasurementsFlowcurves2001}.

The flow dynamics proceeds through irreversible plastic rearrangements that are localized both in space and time and generate long-ranged elastic stress redistributions~\cite{princenRheologyFoamsHighly1983,falk_dynamics_1998,maloneyAmorphousSystemsAthermal2006,tanguyPlasticResponse2D2006,schallStructuralRearrangementsThat2007,lernerScalingTheorySteadystate2009} [Fig.~\ref{fig:resume}(e)].
Elasto-plastic models~\cite{nicolas_deformation_2018}, based on this mesoscopic description, provide minimal statistical-physics models for the yielding transition.
They quantitatively reproduce the avalanche statistics associated with plastic deformation~\cite{talamali2011avalanches,lin_criticality_2015,liuDrivingRateDependence2016,budrikis2017universal,ferreroCriticalityElastoplasticModels2019}, while also capturing the qualitative rheological behavior of amorphous materials~\cite{picardSlowFlowsYield2005,martensSpontaneousFormationPermanent2012a,nicolasRheologyAthermalAmorphous2014}.

Within the elasto-plastic framework, the yielding transition under imposed shear stress $\Sigma$ (rather than imposed shear rate) can also be interpreted as an APT~\cite{nicolas_deformation_2018}. 
The active phase occurs for $\Sigma>\Sigma_Y$, where the plastic activity $A$, proportional to the shear rate $\dot\gamma$, follows the scaling relation
$A \sim (\Sigma-\Sigma_Y)^\beta$,
with $\beta = 1/n$ [Fig.~\ref{fig:resume}(f)].
By contrast, below the yield stress, plastic deformation ceases after a finite accumulated strain ($A=0$), leaving the system arrested in a jammed absorbing state.
For yielding too, there is a conserved quantity (the mechanical stress) coupled to the activity, and an infinite number of absorbing states (all microscopic configurations in mechanical equilibrium for a given applied stress below the yield value).

While the yielding transition and the RIT exhibit a depinning-like  phenomenology at a qualitative level, the presence of long-range interactions leads to critical behaviors near the APT which differ from the CDP class~\cite{lin_scaling_2014,lin_criticality_2015,tyukodiDepinningTransitionPlastic2016,ferreroElasticInterfacesDisordered2019,jocteurYieldingAbsorbingPhase2024a,mariAbsorbingPhaseTransitions2022,jocteurRandom2025}.
They are convex transitions ($\beta>1$), with vanishing order parameter fluctuations, as well as normal large-scale density fluctuations, which are non-standard from the point of view of usual APTs in the CDP class which display hyperuniformity (i.e., anomalously low density fluctuations on large scales)
at the critical point~\cite{hexnerHyperuniformityCriticalAbsorbing2015,tjhungHyperuniformDensityFluctuations2015,leiHydrodynamicsRandomorganizingHyperuniform2019,maTheoryHyperuniformityAbsorbing2023,maHyperuniformityAbsorbingState2025,wieseHyperuniformityMannaModel2024}.
In both cases, these properties have been argued~ \cite{hebraudModeCouplingTheoryPasty1998,bocquetKinetic2009,nicolas_deformation_2018,mariAbsorbingPhaseTransitions2022,jocteurRandom2025}
to stem from an original mechanism, namely that the activity is induced by mechanical noise which arises from incoherent contributions from active regions mediated by long-range interactions, of hydrodynamic or elastic nature, decaying spatially as $1/r^\alpha$. 
This mechanism, inherently tied to long-range interactions, differs from the long-range hopping found in other systems showing APTs~\cite{hinrichsenModelAnomalousDirected1999,ginelliDirectedPercolationLongrange2005,ginelliContactProcessesLong2006,hinrichsenNonequilibriumPhaseTransitions2007,argoloVanishingOrderparameterCritical2013}, or from thermal-like noise reactivation that leads to thermal rounding and thermal avalanches in depinning-like problems~\cite{chenInterfaceMotionRandom1995,bustingorryThermalRoundingDepinning2007,hexnerEnhancedHyperuniformityRandom2017,ferreroCreepMotionElastic2021,korchinskiThermalAvalanchesDrive2025}.
Moreover, the same minimal model (the so-called Hébraud-Lequeux (HL) model \cite{hebraudModeCouplingTheoryPasty1998}) has been argued to provide an approximate mean-field description of both the yielding transition and the RIT~\cite{hebraudModeCouplingTheoryPasty1998,agoritsasRelevanceDisorderAthermal2015,linMeanFieldDescriptionPlastic2016,linMicroscopicProcessesControlling2018,jocteurRandom2025}. 

The shared atypical critical properties and the qualitatively similar  microscopic mechanism for activity creation beg for further exploration of the similarities between these two transitions, with the underlying idea that the yielding transition and the RIT may actually be two physical realizations of the same transition. 
This is the question we address in this article. 
Because the critical behaviors of both transitions have already been characterized in the literature, our contribution is not primarily regarding these properties (although we also contribute some original results on spatial correlations and avalanches near a short-range analogue of the yielding transition), but rather regarding the comparison between the two transitions.
Because the way long-range interactions generate mechanical noise differs between the two transitions, for a given value of the interaction exponent $\alpha$, the two transitions are not readily comparable.
We therefore compare the evolution of the criticality of generalized models of yielding and RIT where the range of interactions $\alpha$ is taken as a free parameter, which allows to make a meaningful comparison in a parametric representation free of $\alpha$.
We also perform a similar analysis for the Hébraud-Lequeux model.
This reveals that all distinctive features of the transition (convexity, vanishing of fluctuations, normal long-range density fluctuations, sparse avalanches) fade away when decreasing the range of interactions, a trend also exhibited by the HL model. 
Remarkably, these features present at low enough $\alpha$ disappear at a unique interaction range $\alpha^\ast$, whose value however differs for the yielding and reversible-irreversible transitions.
For $\alpha>\alpha^\ast$, we argue that both yielding and RIT are effectively described by the long-range CDP class (with an effective power-law exponent $\alpha'$), which encompasses long-range transport of a conserved quantity.

The paper is organized as follows. In Sec.~\ref{Sec:def:models}, we recall the definitions of the Random Organization Model (ROM) used to model the RIT, and of Picard's elastoplastic model for the yielding transition, together with the generalizations of these models including tunable long-range interactions.
Then, in Sec.~\ref{sec:similar:crit} we compare the static and dynamic critical properties of both models as a function of the decay exponent $\alpha$ of their long-range mediated interactions, highlighting similarities and differences.
Finally, in Sec.~\ref{sec:theoretical:frameworks} we discuss the relevance of two theoretical frameworks to describe these transitions, the long-range CDP field equations describing long-range transport, as well as the Hébraud-Lequeux model describing the effect of long-range mediated noise, to account for the critical behaviors of the ROM and Picard models with tunable long-range interactions.


\section{Definition of the models}
\label{Sec:def:models}

In this section, we define the two models considered in this paper, which both display an absorbing phase transition in the presence of long-range interactions.
The first model is a generalization, including long-range mediated interactions, of the Random Organization Model describing the stroboscopic dynamics of cyclically-sheared suspensions introduced in \cite{jocteurRandom2025}. The second one is the Picard elastoplastic model describing the mesoscale dynamics of a yield-stress fluid \cite{picardSlowFlowsYield2005}.

\subsection{Generalized Random Organization Model for cyclically-sheared suspensions}
\label{sec:def:ROM}

As mentioned in the introduction, cyclically-sheared suspensions exhibit an absorbing phase transition as a function of either the oscillatory shear amplitude, or the density of suspended particles. As often with phase transitions, the behavior close to the critical point may be described using minimal models retaining only key ingredients.
Along this line, the absorbing phase transition in cyclically-sheared suspensions has been modeled using a simple interacting particle model called Random Organization Model (ROM) \cite{corteRandomOrganizationPeriodically2008,corteSelfOrganizedCriticalitySheared2009,schrenkCommunicationEvidenceNonergodicity2015,tjhungHyperuniformDensityFluctuations2015,tjhungCriticalityCorrelatedDynamics2016}.
This model exhibits a transition which belongs to the Conserved Directed Percolation (CDP) universality class~\cite{hexnerHyperuniformityCriticalAbsorbing2015}.
However, it has recently been shown that taking into account hydrodynamic long-range interactions between particles in the suspension significantly modifies the critical properties of the transition, which becomes convex ($\beta>1$) \cite{mariAbsorbingPhaseTransitions2022,jocteurRandom2025}.
In the following subsections, we start by recalling the definition of the ROM, before moving to its generalized version with long-range mediated interactions, called $\alpha$-ROM, which will be considered in the rest of the paper.

\subsubsection{Random Organization Model}

We first introduce the two-dimensional version of the original ROM, which is composed of $\Ntot$ discs of diameter $D$ located at position $\bm{r}^t_i$ ($i=1,\dots,\Ntot$) at time $t$,
within a square box of linear size $L$, subject to periodic boundary conditions.
We define the distance $r^t_{ij}$ between particles $i$ and $j$ as $r^t_{ij} = |\bm{r}^t_i -\bm{r}^t_j|$, taking into account periodic boundary conditions (in other words, $r^t_{ij}$ is the shortest distance between particles $i$ and $j$).
Overlapping particles, whose centers are closer than their diameter $D$ (i.e., $r^t_{ij}<D$) are called active, while other particles are said to be passive.
The physical interpretation of a pair of overlapping particles in relation to experiments on cyclically-sheared suspensions is that the two particles make a contact during a cycle, which alters their trajectories. 
In this spirit, the diameter is rather to be interpreted as an effective diameter related to the amplitude of the shear cycle (and not the physical diameter of particles).
The ROM dynamics is defined at discrete time, in analogy with the stroboscopic observation in oscillatory shear experiments.
In a time step from $t$ to $t+1$, particles positions evolve as
\begin{equation}
    \bm{r}^{t+1}_i = \bm{r}^{t}_i + \bm{\delta}_{a,i}^t\, .
\end{equation}
where $\bm{\delta}_{a,i}^t$ is a random kick with typical size $\Delta_a = D$.
These random displacements model particles motion due to contacts during a cycle in an experiment.
At each discrete time $t$, particles are labelled as active or passive according to whether they overlap or not with another (or several) particles.
If only passive particles are present in the system, no more dynamics takes place, which corresponds to an absorbing state.

Varying particle area fraction $\phi = \Ntot \pi (D/2)^2/L^2$, an absorbing phase transition occurs at a critical value $\phi_c$, which thus plays the role of the key control parameter of the transition \cite{tjhungHyperuniformDensityFluctuations2015,tjhungCriticalityCorrelatedDynamics2016}. 
For $\phi<\phi_c$, the system falls at large time into an absorbing phase, while for $\phi>\phi_c$, it remains in a statistically active state, where the fraction $A$ of active particles remains non zero and fluctuates around an average value $\langle A \rangle>0$, which plays the role of the order parameter of the absorbing phase transition.

\subsubsection{Random Organization Model with long-range mediated interactions ($\alpha$-ROM)}

Importantly, the simplest version of the ROM, defined above, neglects hydrodynamic interactions between particles mediated by the surrounding fluid.
However, long-range mediated interaction have been shown to have a strong impact on the critical properties of ROM-type models
\cite{weijsEmergentHyperuniformityPeriodically2015,mariAbsorbingPhaseTransitions2022,jocteurRandom2025}.
We consider here the simple implementation of long-range mediated interactions proposed in \cite{jocteurRandom2025},
where passive particles are subject to a small random displacement of amplitude $\sim 1/r^{\alpha}$ ($\alpha>0$) induced by active particles located at a distance $r$ from the passive particle. Displacements induced by different active particles are additive, resulting in a Gaussian total displacement for the passive particle.
We denote this generalized version of Random Organization Model with long-range mediated interactions as the $\alpha$-ROM, to emphasize the role of the power-law interaction decay exponent $\alpha$ on the behavior of the model.

For an efficient numerical implementation, we evaluate the variance of the displacement of the passive particle from an activity field coarse-grained over boxes of linear size $D$.
For box $b$, the activity is defined as 
$A_b = \mathcal{N}_b^{\mathrm{act}}/D^2$, where $\mathcal{N}_b^{\mathrm{act}}$ is the number of active particles whose center position is within the box.
At a given time step $t$, all passive particles are moved by a random displacement vector $\bm{\delta}_{p,i}^t$. Each component of $\bm{\delta}_{p,i}^t$ has a Gaussian
distribution with zero mean and variance $\Delta_{\mathrm{p},i}^2$ evaluated as
\begin{equation}
    \Delta_{\mathrm{p},i}^2 = \sum_{b'} G(r_{b'b}) A_{b'}\, ,
    \label{eq:passive_step_size}
\end{equation}
where $r_{b'b}$ is the center-to-center distance between boxes $b'$ and $b$ (where the active and passive particles are respectively located), and $G(r)$ is the propagator
\begin{equation}
    G(r) = \frac{c}{(1+r^2)^{\alpha}}.
    \label{eq:kernel}
\end{equation}
A more detailed discussion of the technical implementation of long-range interactions and of the underlying physical assumptions can be found in \cite{jocteurRandom2025}.

Similarly to the ROM, the $\alpha$-ROM also displays an absorbing phase transition, but at a shifted critical packing fraction $\phi_{c,\alpha}$, due to the presence of long-range mediated interactions. The critical properties of the $\alpha$-ROM are discussed in Sec.~\ref{sec:similar:crit}.


\subsection{Elastoplastic model for the flow of yield-stress fluids}
\label{sec:def:Picard}

We now turn to a second class of soft matter systems in which long-range interactions play a key role, namely yield-stress fluids.
These are complex fluids which flow only if the applied shear stress is larger than a threshold value called yield stress.
In the elastoplastic scenario \cite{nicolas_deformation_2018}, plastic deformation is spatially heterogeneous and occurs through localized plastic relaxation events, during which shear stress is redistributed in the surrounding elastic matrix through a long-range elastic interaction kernel \cite{tanguyPlasticResponse2D2006,nicolas_deformation_2018}. 
We describe below a standard mesoscopic elastoplastic model, the Picard model, characterizing the flow of yield-stress fluids, before introducing a generalization of this model with tunable long-range interactions.

\subsubsection{Picard elastoplastic model}

The Picard elastoplastic model~\cite{picardSlowFlowsYield2005} is a simple dynamical model describing stress redistribution in the elastoplastic scenario for the flow of yield stress fluids. 
The model is defined on a two dimensional square lattice of linear extension $L$, with $\Ntot=L^2$ sites labeled by the indices $(i,j)$.
Mechanical variables like stress or strain defined on each lattice site may be interpreted as mesoscale, fluctuating continuum quantities. 
More specifically, each site carries several local mechanical variables: a shear stress $\sigma_{i,j}$, a plastic strain $\epsilon_{i,j}$,
and a mechanical state variable $n_{i,j}$ defined as $n_{i,j} = 1$ if the site is in a plastic state, or $n_{i,j} = 0$ if elastic.
Note that the Picard model is a scalar stress model, taking into account only the shear stress component of the stress tensor.
The time evolution of these variables can be expressed as:
\begin{equation}
    \partial_t\sigma_{i,j} = g_0 \sum_{k,l} G^\mathrm{E}_{i-k, j-l}\, \dot{\epsilon}_{k,l}\, , 
    \qquad \dot{\epsilon}_{i,j} = \frac{n_{i,j}\sigma_{i,j}}{g_0 \tau}\, ,
    \label{eq:picard_model_1}
\end{equation}
and
\begin{equation}
    n_{i,j}:  \begin{cases}
            0 \xrightarrow{\tau^{-1}} 1 \,, & \text{if } |\sigma_{i,j}|>\sigma_\mathrm{c} \\
            1 \xrightarrow{\tau^{-1}} 0 \,, & \forall \sigma_{i,j}
            \end{cases} 
\label{eq:picard_model_3}
\end{equation}
where $g_0=1$ is the shear elastic modulus, $\tau=1$ is the elastic relaxation time, and $\sigma_\mathrm{c}$ is a site-independent threshold for local yielding, meaning for the local transition from the elastic to plastic state.
When a site $(i,j)$ becomes plastic, its local stress $\sigma_{i,j}$ is immediately released and redistributed according to the discretized Eshelby kernel~\cite{eshelbyDeterminationElasticField1957,picardElasticConsequencesSingle2004}, defined via its Fourier transform
\begin{equation} \label{eq:Eshelby}
    \tilde{G}^\mathrm{E}_{q_x, q_y} =  -\frac{4 q_x^2 q_y^2}{|\bm{q}|^4} \qquad \text{for } \bm{q} \equiv (q_x, q_y) \neq (0, 0),
\end{equation}
and $\tilde{G}^\mathrm{E}_{0,0}=0$.
In the present model, the total stress $\Sigma=\overline{\sigma_{i,j}}$ is conserved by the dynamics, and the fixed value $\Sigma$ is thus used as the control parameter, called the applied stress.
The activity (here the flowing state) is quantified by the shear rate $\dot\gamma = \overline{\dot{\epsilon}_{i,j}}$, which is determined as an outcome of the dynamics.
The Eshelby propagator considered here satisfies mechanical equilibrium under constant applied stress, which translates into the conditions
$\sum_{k'} G^\mathrm{E}_{k', l} = \sum_{l'} G^\mathrm{E}_{k, l'} = 0$ for all $(k,l)$, valid since $\tilde{G}^\mathrm{E}_{q_x,0} = \tilde{G}^\mathrm{E}_{0,q_y} = 0$ as seen from Eq.~(\ref{eq:Eshelby}).
This property is sometimes referred to as the presence of soft modes, or zero modes \cite{linDensityShearTransformations2014,lin_scaling_2014,tyukodiDepinningTransitionPlastic2016,ferreroElasticInterfacesDisordered2019,jocteurYieldingAbsorbingPhase2024a}.
This implies in particular that the propagator cannot have a uniform sign. Indeed, the Eshelby propagator has a four-fold angular symmetry $\propto \cos 4\theta$
in a polar representation $\bm{r}=(r,\theta)$.

The Picard model displays a yielding transition at a critical value $\Sigma_Y$ of the applied stress $\Sigma>0$. For $\Sigma<\Sigma_Y$, no flow occurs ($\dot\gamma=0$),
while for $\Sigma>\Sigma_Y$, a stationary flow sets in ($\dot\gamma>0$) \cite{nicolas_deformation_2018}.
The yielding transition has been characterized as an absorbing phase transition \cite{jocteurYieldingAbsorbingPhase2024a}.
The yield stress $\Sigma_Y$ therefore plays the role of a critical stress value.

\subsubsection{Tunable long-range interactions: the $\alpha$-Picard model}

The Eshelby propagator falls off at large distance $r$ as $1/r^{d}$, where the exponent $\alpha=d$ is constrained by linear elasticity ($d$ being the space dimension, here $d=2$).
However, in a statistical physics perspective, it may be of interest to consider lattice elastoplastic models not only as models describing the mechanical state of complex fluids,
but also as more abstract minimal models to study the critical properties of absorbing phase transitions in the presence of long-range interactions.
In this view, one may rather consider the exponent $\alpha$ as a tunable exponent which controls the critical behavior, rather than a exponent fixed to $\alpha=d$ by mechanical constraints.

With this goal in mind, we thus need to properly generalize the Eshelby propagator to the case $\alpha \neq 2$, denoted here as $G^{\mathrm{E},\alpha}_{k,l}$
(we focus here exclusively on the two-dimensional case). We follow here the definition of the generalized Eshelby propagator given in \cite{jocteurYieldingAbsorbingPhase2024a}.
For consistency with the standard Eshelby propagator, we aim at keeping the presence of zero modes, which in discretized Fourier space reads as
$\tilde{G}^{\mathrm{E},\alpha}_{q_x, 0} =  \tilde{G}^{\mathrm{E},\alpha}_{0,q_y} = 0$.
It is thus convenient to define the propagator $G^{\mathrm{E},\alpha}_{k,l}$ by its Fourier-space representation $\tilde{G}^{\mathrm{E},\alpha}_{q_x, q_y}$ as
\begin{equation} \label{eq:Eshelby:alpha}
    \tilde{G}^{\mathrm{E},\alpha}_{q_x, q_y} = -b_\alpha \, \frac{q_x^2 q_y^2}{|\bm{q}|^{6-\alpha}} \qquad \text{for } \bm{q} \equiv (q_x, q_y) \neq (0, 0),
\end{equation}
and $\tilde{G}^{\mathrm{E},\alpha}_{0,0}=0$.
The real space propagator $G^{\mathrm{E},\alpha}_{k,l}$ can then be expressed in a continuum limit $L/a \to \infty$ (where $a$ is the lattice spacing) in the scaling form
\begin{equation}
G^{\mathrm{E},\alpha}_{k,l} = (a/L)^\alpha \mathcal{G}_\alpha(\bm{r}),
\end{equation}
with $\bm{r} = (ak/L,al/L)$. The rescaled propagator $\mathcal{G}_\alpha(\bm{r})$ is given in polar form 
$\bm{r} = (r\cos\theta, r\sin\theta)$
\begin{equation}
\mathcal{G}_\alpha(\bm{r}) \propto [C_\alpha + \cos 4\theta]/r^{\alpha}
\end{equation}
where $C_\alpha$ is a constant. 
In the physical case of the yielding transition, corresponding to $\alpha = 2$ in two dimensions, one has $C_2 = 0$.
The dynamics of this generalized Picard model with the propagator $G^{\mathrm{E},\alpha}_{k,l}$ follows the same dynamics as the standard Picard model
[see Eqs.~(\ref{eq:picard_model_1}) and (\ref{eq:picard_model_3})], upon replacement of the standard Eshelby propagator $G^{\mathrm{E}}_{k,l}$ by $G^{\mathrm{E},\alpha}_{k,l}$.
We call the resulting model the $\alpha$-Picard model \cite{jocteurYieldingAbsorbingPhase2024a}.

The $\alpha$-Picard model also displays an absorbing phase transition at a value $\Sigma_{Y,\alpha}$ of the applied stress $\Sigma$.
The critical properties of this model are discussed below.


\section{A qualitatively similar evolution of the critical behavior}
\label{sec:similar:crit}

Having defined both the $\alpha$-ROM and $\alpha$-Picard models, we now discuss and compare their critical properties, with the aim of highlighting their similarities and differences.
In both models, critical exponents are defined in terms of the scaling of observables with the distance $\varepsilon$ to the critical point:
$\varepsilon=(\phi-\phi_{c,\alpha})/\phi_{c,\alpha}$ in the $\alpha$-ROM, and $\varepsilon=(\Sigma-\Sigma_{Y,\alpha})/\Sigma_{Y,\alpha}$ in the $\alpha$-Picard model.
To make comparison easier, we generically denote as $A$ the activity in both models. In the $\alpha$-ROM, it corresponds to the fraction $A$ of overlapping particles as defined in Sec.~\ref{sec:def:ROM}, while in the $\alpha$-Picard model, it can be chosen as the shear rate, $A=\dot\gamma$, see Sec.~\ref{sec:def:Picard}.

\subsection{From concave to convex transitions when increasing the interaction range}

\begin{figure}[h]
    \centering
    \includegraphics[width=0.6\linewidth]{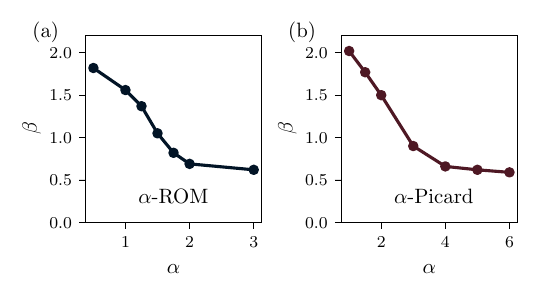}
    \caption{Critical exponent $\beta$ characterizing the mean activity $\langle A \rangle \sim \varepsilon^{\beta}$ as a function of the power-law decay exponent $\alpha$ of the propagator, where $\varepsilon$ is the distance to the critical point. (a) $\alpha$-ROM, with $\varepsilon=(\phi-\phi_{c,\alpha})/\phi_{c,\alpha}$. Data originally from~\cite{jocteurRandom2025}. (b) $\alpha$-Picard model, with  $\varepsilon=(\Sigma-\Sigma_{Y,\alpha})/\Sigma_{Y,\alpha}$. Data originally from~\cite{jocteurYieldingAbsorbingPhase2024a}.}
    \label{fig:beta_vs_alpha}
\end{figure}

The critical behavior of the $\alpha$-ROM depends on the exponent $\alpha$ characterizing long-range interactions, as shown in Fig.~\ref{fig:beta_vs_alpha}(a) for the critical exponent $\beta$ characterizing the mean activity $\langle A \rangle \sim \varepsilon^{\beta}$. For $\alpha \gtrsim 3$, the propagator decays rapidly enough, and the short range-behavior $\beta \approx 0.64$ is recovered \cite{lubeckUniversalScalingBehavior2004}.
In contrast, for $\alpha \lesssim 0.5$, the propagator decays very slowly and the infinite-range behavior $\beta \approx 1.85$ is obtained \cite{mariAbsorbingPhaseTransitions2022}. 
In the intermediate range, $\beta$ continuously depends on $\alpha$, see Fig.~\ref{fig:beta_vs_alpha}(a) \cite{jocteurRandom2025}.
Note that a continuous dependence on the power-law decay exponent $\alpha$ also exists for equilibrium systems with long-range interactions \cite{fisherCriticalExponentsLongRange1972,hinrichsenNonequilibriumPhaseTransitions2007}.
Here, an interesting phenomenon is that long-range interactions do not only push the value of $\beta$ towards its mean-field CDP value $\beta_{MF}=1$, but $\beta$ further increases beyond $\beta_{MF}$, and reaches values up to $\beta \approx 2$. The absorbing phase transition thus turns from concave ($\beta <1$) to convex ($\beta >1$). The crossing point at which $\beta=\beta_{MF}=1$ is around $\alpha \approx \frac{3}{2}$ [Fig.~\ref{fig:beta_vs_alpha}(a)].

A qualitatively similar behavior is observed for the $\alpha$-Picard model as seen in Fig.~\ref{fig:beta_vs_alpha}(b) \cite{jocteurYieldingAbsorbingPhase2024a}, but with shifted threshold values of the decay exponent $\alpha$ with respect to the $\alpha$-ROM case.
For the $\alpha$-Picard model, the mean-field regime $\beta=2$ (corresponding to the H\'ebraud-Lequeux model \cite{hebraudModeCouplingTheoryPasty1998}, see Sec.\ref{sec:theoretical:frameworks}) is observed for $\alpha \lesssim 1$. For $1 \lesssim \alpha \le 6$, the exponent $\beta$ continuously depends on $\alpha$, with a weaker $\alpha$-dependence for $\alpha>4$ \cite{jocteurYieldingAbsorbingPhase2024a}.
Indeed, a careful analysis reveals that the presence of zero modes may slightly alter the value of critical exponent for $4<\alpha<6$ in the $\alpha$-Picard model, leading to $\beta \approx 0.59$ for $\alpha >6$ \cite{jocteurYieldingAbsorbingPhase2024a}.
We show in Sec.~\ref{sec:hyperuniformity} that the effect of zero modes is more dramatic for spatial correlations of the order parameter. 
As for the $\alpha$-ROM, a crossover from a concave to a convex transition takes place, here around $\alpha \approx 3$.

\subsection{From diverging to vanishing critical fluctuations}

\begin{figure}[h]
    \centering
    \includegraphics[width=0.6\linewidth]{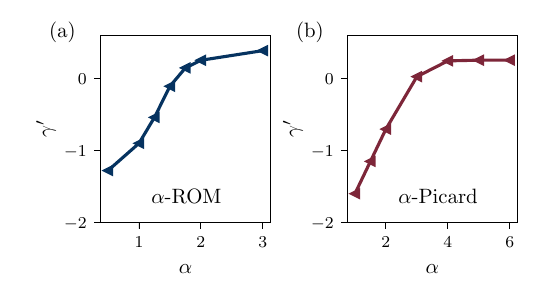}
    \caption{Critical exponent $\gamma'$ characterizing the variance of the activity $\langle (A-\langle A \rangle)^2 \rangle \sim \varepsilon^{-\gamma'}$ as a function of the power-law decay exponent $\alpha$ of the propagator. (a) $\alpha$-ROM, with $\varepsilon=(\phi-\phi_{c,\alpha})/\phi_{c,\alpha}$. Data originally from~\cite{jocteurRandom2025}. (b) $\alpha$-Picard model, with  $\varepsilon=(\Sigma-\Sigma_{Y,\alpha})/\Sigma_{Y,\alpha}$. Data originally from~\cite{jocteurYieldingAbsorbingPhase2024a}.}
    \label{fig:gamma_vs_alpha}
\end{figure}

Beyond the mean value of the order parameter, the variance of its fluctuations can also be measured, leading to the critical exponent $\gamma'$ defined as
$\langle (A-\langle A \rangle)^2 \rangle \sim \varepsilon^{-\gamma'}$. The minus sign in the definition of $\gamma'$ is conventional, and comes from the fact that in most cases fluctuations of the order parameter diverge at the critical point.
The value of $\gamma'$ is plotted as a function of $\alpha$ for the $\alpha$-ROM in Fig.~\ref{fig:gamma_vs_alpha}(a) \cite{jocteurRandom2025}
and the $\alpha$-Picard model in Fig.~\ref{fig:gamma_vs_alpha}(b) \cite{jocteurYieldingAbsorbingPhase2024a}.
Both curves show a similar trend, with a decreasing value of $\gamma'$ when $\alpha$ decreases, that is, when the interaction range is increased \cite{jocteurYieldingAbsorbingPhase2024a,jocteurRandom2025}.
The range of $\alpha$ over which $\gamma'$ continuously depends on $\alpha$ is the same, up to numerical accuracy, as the one reported for the $\beta$ exponent in each model.
Interestingly, the value of $\gamma'$ becomes negative for low values of $\alpha$, meaning that in this regime activity fluctuations vanish instead of diverging as in usual critical phenomena. However, this behavior does not contradict the divergence of the correlation length, since the variance of activity fluctuations still diverges with respect to the mean value $\langle A \rangle$ of activity, i.e., $\langle (A-\langle A \rangle)^2 \rangle / \langle A \rangle^2$ diverges at the critical point (equivalently, $\gamma'+2\beta>0$).
The crossing point $\gamma'=0$ is found in each model at values of $\alpha$ which are very close to the crossover values between concave and convex transitions.

Interestingly, the relationship between the convex character of the transition and the associated vanishing of fluctuations can be rationalized, at a qualitative level,
through the hyperscaling relation. The latter relates the exponents $\gamma'$ and $\beta$ to the exponent $\nu_\perp$ characterizing the critical divergence of the correlation length
$\xi \sim \varepsilon^{-\nu_\perp}$ as
\begin{equation}
    \gamma^\prime = d\nu_\perp - 2\beta ,
\end{equation}
with $d$ the space dimension \cite{fisher1973general,fisherRenormalizationGroupTheory1974,lubeckUniversalScalingBehavior2004}.
Note that the hyperscaling relation has been verified numerically in the 2D Picard model (with $\alpha=2$, corresponding to the physical Eshelby propagator), and that it can be interpreted in a scaling scenario
in terms of the largest avalanches dominating the creation of activity in the system \cite{lin_scaling_2014}.
Assuming that $\nu_\perp$ has a relatively weak dependence on $\alpha$, we see that increasing $\beta$ leads to a decrease of $\gamma'$, consistently with the general trend reported in both the $\alpha$-ROM and $\alpha$-Picard models. In addition, in the mean-field CDP case resulting from long-range interactions, one has $\beta=1$ and $\nu_\perp=1$
\cite{hinrichsenNonequilibriumPhaseTransitions2007}, resulting in the two-dimensional case in $\gamma'=0$ according to the hyperscaling relation. Hence in 2D, the onset of a convex transition is accompanied by the onset of vanishing fluctuations.
Note that the vanishing of critical fluctuations has also previously been reported in the context of convex absorbing phase transitions 
\cite{argoloVanishingOrderparameterCritical2013,villarroelCriticalYieldingRheology2021a,mariAbsorbingPhaseTransitions2022}.

\subsection{Loss of avalanches compacity}

\begin{figure}[t]
    \centering
    \includegraphics[width=0.6\linewidth]{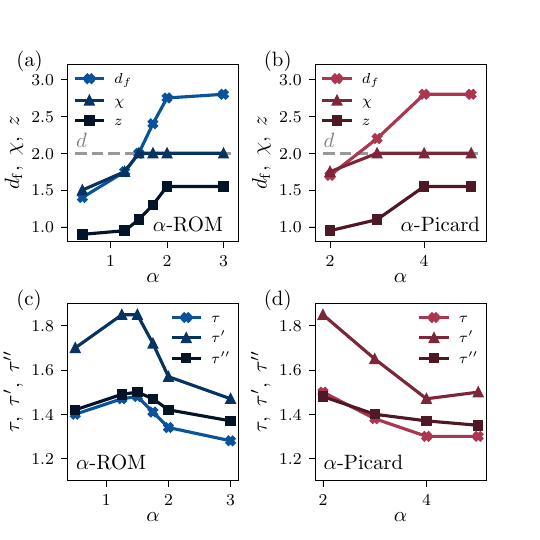}
    \caption{Critical exponents of avalanche statistics versus the power-law decay exponent $\alpha$ of the propagator. Top: fractal dimension $d_f$, exponent $\chi$ and dynamical exponent $z$ characterizing the cut-off of avalanche distributions, for the $\alpha$-ROM model (a) (data originally from~\cite{jocteurAvalanches2026}) and the $\alpha$-Picard model (b) (data originally from~\cite{jocteurProtocol2025}). Bottom: power-law exponents $\tau$, $\tau'$ and $\tau''$ for the $\alpha$-ROM model (c) (data originally from~\cite{jocteurAvalanches2026}) and the $\alpha$-Picard model (d) (data originally from~\cite{jocteurProtocol2025} for $\alpha=2$, original data otherwise).}
    \label{fig:avalanche_vs_alpha}
\end{figure}

At the critical point, both the $\alpha$-ROM and $\alpha$-Picard models display an intermittent dynamics, in the form of avalanches of correlated events.
This is a generic property of the dynamics of a system at the critical point of an absorbing phase transition.
We denote as $S$ the avalanche size (i.e., the number of events taking place in a given avalanche), $T$ the avalanche duration, and $N$ the number of particles (for the $\alpha$-ROM) or sites (for the $\alpha$-Picard model) involved in the avalanche. By definition, $N \le \Ntot$, where $\Ntot$ is the total number of particles or sites, whereas the size $S$ and duration $T$ generically have no strict upper bound, although the finite size of the system does induce an upper characteristic value.

As expected at a critical point, the quantities $S$, $T$ and $N$ exhibit a power-law behavior, with an upper (soft) cut-off depending on system size in a power-law manner.
We write the corresponding probability density functions as
\begin{align}
    P_S(S) & \sim S^{-\tau} g_S\left(\frac{S}{S_\mathrm{c}}\right), \quad S_\mathrm{c}\sim L^{d_f} \\
    P_T(T) & \sim T^{-\tau'} g_T\left(\frac{T}{T_\mathrm{c}}\right), \quad T_\mathrm{c}\sim L^z \\
    P_N(N) & \sim N^{-\tau''} g_N\left(\frac{N}{N_\mathrm{c}}\right), \quad N_\mathrm{c}\sim L^\chi
\end{align}
which defines three power-law exponents $\tau$, $\tau'$ and $\tau''$, as well as three exponents $d_f$, $z$ and $\chi$ for the scaling of the upper cut-offs with the linear system size $L$. The exponent $d_f$ is called the fractal dimension of avalanches, while the exponent $z$ is called the dynamical exponent as it relates by a scaling law temporal and spatial properties.
The scaling exponents $d_f$, $z$ and $\chi$, as well as the power-law exponents $\tau$, $\tau'$ and $\tau''$, are plotted as a function of $\alpha$ in Fig.~\ref{fig:avalanche_vs_alpha}
for the $\alpha$-ROM in panels (a,c) \cite{jocteurAvalanches2026} and the $\alpha$-Picard model in panels (b,d) \cite{jocteurProtocol2025}.


\subsection{A highlight on similarities in the evolution of critical exponents}

\subsubsection{Coinciding changes of behaviors with $\alpha$}

\begin{figure}[h]
    \centering
    \includegraphics[width=0.6\linewidth]{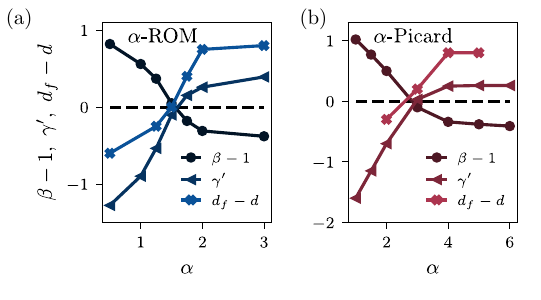}
    \caption{Plots of $\beta-1$, $\gamma'$ and $d_f-d$ (data from Figs.~\ref{fig:beta_vs_alpha}, \ref{fig:gamma_vs_alpha} and \ref{fig:avalanche_vs_alpha}) showing that the values $\beta=1$, $\gamma'=0$ and $d_f=d$ (here, $d=2$) occur for very close values of $\alpha \approx \alpha^\ast$, suggesting that the common value $\alpha^\ast$ separates two markedly different behaviors. (a) $\alpha$-ROM, (b) $\alpha$-Picard model.}
    \label{fig:parallel_evolution}
\end{figure}

When discussing in the above sections the evolution of critical exponents with the decay exponent $\alpha$ of the propagator, we have observed in both the $\alpha$-ROM and $\alpha$-Picard models three different changes of behavior when decreasing $\alpha$: (i) a change from a concave ($\beta<1$) to a convex ($\beta>1$) transition;
(ii) a change from diverging ($\gamma'>0$) to vanishing ($\gamma'<0$) fluctuations; (iii) a change from compact ($d_f>d$) to non-compact ($d_f<d$) avalanches.
Interestingly, these three transitions occur around the same (model-dependent) value of $\alpha$, as seen in Fig.~\ref{fig:parallel_evolution} by plotting $\beta-1$, $\gamma'$ and $d_f-d$ as a function of $\alpha$ on the same panel for each model, to show that the three quantities all vanish for very close values of $\alpha$, say $\alpha \approx \alpha^\ast$.

The critical state obtained for $\alpha = \alpha^\ast$ has a mean-field flavour, as the critical exponent values $\beta=1$, $\gamma'=0$ and $d_f=d$ correspond to the mean-field CDP values
\cite{lubeckUniversalScalingBehavior2004}.
However, it does not share all the characteristics of the mean-field CDP behavior. For instance, the critical decay $A(t) \sim t^{-\delta}$ obtained at the critical point ($\varepsilon=0$) is characterized by $\delta \approx 0.5$ in the $\alpha$-ROM in 2D \cite{jocteurRandom2025}, while the corresponding mean-field CDP value is $\delta_{MF}=1$.

\subsection{Parametric plots of critical exponents}

\begin{figure}[t]
    \centering
    \includegraphics[width=0.5\textwidth]{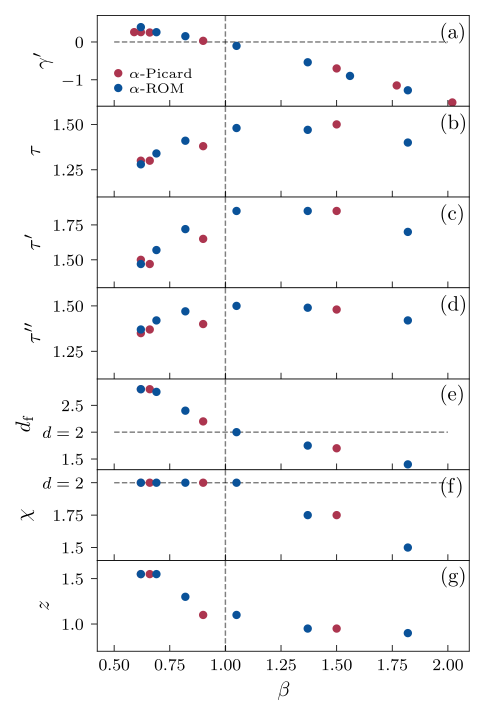}
    \caption{Parametric plots of the exponents as a function of the exponent $\beta$, showing a rather good collapse of data from different models. (a)  $\gamma'$, (b) $\tau$, (c)  $\tau'$, (d)  $\tau''$, (e)  $d_f$, (f)  $\chi$, (g) $z$.}
    \label{fig:parametric_exponents}
\end{figure}

We have just seen that the change of behavior from a concave to a convex transition, based on the critical exponent $\beta$, also reflects on the other critical exponents
$\gamma'$ and $d_f$ which characterize different critical properties. This corresponds to a global change of critical properties when crossing $\alpha=\alpha^\ast$.
However, the value $\alpha^\ast$ is model-dependent: $\alpha^\ast \approx 1.5$ in the $\alpha$-ROM, whereas $\alpha^\ast \approx 3$ in the $\alpha$-Picard model.
Hence a quantitative comparison of critical exponents plotted versus $\alpha$ for the $\alpha$-ROM and $\alpha$-Picard models is a priori bound to fail.
This discrepancy may be physically understood in the following way.
In the $\alpha$-Picard model, the conserved (shear) stress $\sigma$ is redistributed over large distances by the Eshelby long-range propagator.
One thus has long-range transport of a conserved quantity which is coupled to the creation of activity, as a locally larger stress $\sigma$ statistically leads to a higher creation of activity. The situation is different in the $\alpha$-ROM, since the conserved field coupled to activity creation is the particle density, which is subject only to short-range transport. Here, the long-range propagator only induces local moves of distant particles. At variance with the $\alpha$-Picard model, the long-range propagator does not transport a conserved quantity coupled to activity creation. The role of the long-range propagator thus differs in both models, and it is therefore expected that the dependence of the critical exponents on $\alpha$ also differs. 

Yet, the qualitatively similar trends observed for the different critical exponents in both models raises the question of whether these models could be related by a mere reparameterization of $\alpha$. For instance, could the behavior of the $\alpha$-ROM (which lacks transport of a conserved quantity) be interpreted in terms of an effective transport mechanism with an effective propagator characterized by a decay exponent $\alpha_{\text{eff}}$?
To test this idea that both models could behave similarly up to a reparameterization of $\alpha$, we plot on Fig.~\ref{fig:parametric_exponents}
the exponent $\gamma'$, as well as the six avalanche exponents $\tau$, $\tau'$, $\tau''$, $z$, $\chi$, and $d_f$, in a parametric way as a function of the exponent $\beta$.
With this representation, short-range interactions are on the left, at small $\beta$ values, while long-range interactions are on the right, at large $\beta$, but it allows us to get rid of the explicit dependence of the critical exponents on $\alpha$.
If the exponents of both models differed only by a reparameterization of $\alpha$, all parametric curves should coincide.
This is indeed what we observe in Fig.~\ref{fig:parametric_exponents}, although for the power-law exponents $\tau$, $\tau'$ and $\tau''$ the plots show some discrepancies for intermediate values of $\beta$. 
Given the range of values spanned by the exponents, the agreement is still remarkable, 
and suggests that indeed both transitions could share at least to a good approximation the same underlying effective description\footnote{Note that in any case we cannot expect a unified description for $\beta < 0.64$, where zero modes of the propagator affect the behavior of the $\alpha$-Picard model, because these are not present for the $\alpha$-ROM.}.
In fact, we will discuss in section~\ref{sec:theoretical:frameworks} the possibility that the clear distinction at $\beta=1$ between the regime of a concave transition ($\beta<1$), with diverging fluctuations ($\gamma'>0$) and compact avalanches ($d_f>d$) on one side, and the opposite regime of a convex transition ($\beta>1$), with vanishing fluctuations ($\gamma'<0$) and non-compact avalanches ($d_f<d$) on the other side, corresponds to two different underlying frameworks which meet at $\beta=1$. 
Before turning to this though, we will highlight another qualitative change occurring at $\beta=1$ regarding the spatial correlations of stress in the $\alpha$-Picard and density in the $\alpha$-ROM.

\subsection{Loss of hyperuniformity}
\label{sec:hyperuniformity}

We end up the comparison of numerical data from the $\alpha$-ROM and $\alpha$-Picard models by discussing the spatial fluctuations of the conserved field in both models.
Models within the CDP universality class show hyperuniformity at the critical point~\cite{hexnerHyperuniformityCriticalAbsorbing2015,tjhungHyperuniformDensityFluctuations2015,leiHydrodynamicsRandomorganizingHyperuniform2019,maHyperuniformityGeneralizedRandom2019,maTheoryHyperuniformityAbsorbing2023,maHyperuniformityAbsorbingState2025,wieseHyperuniformityMannaModel2024}, that is, large-scale fluctuations of the conserved field are anomalously low~\cite{torquatoLocalDensityFluctuations2003,torquatoHyperuniformStatesMatter2018}. 

For systems of particles like the ROM, which display an absorbing phase transition falling into the CDP class, this translates into a vanishing of the structure factor\footnote{We recall that the structure factor $S(\bm{q})$ is defined in a system of $\Ntot$ particles at positions $\bm{r}_i$ as
\begin{equation}
S(\bm{q}) = \Ntot^{-1} \left\langle \sum_{j,k} e^{\mathrm{i} \bm{q}\cdot(\bm{r}_j - \bm{r}_k)} \right\rangle\, .
\end{equation}
In an isotropic system, $S(\bm{q})$ depends only on $q=|\bm{q}|$, so that it is simply denoted here as $S(q)$.}
$S(q)$ when $q\to 0$.

\begin{figure}[h]
    \centering
    \includegraphics[width=\linewidth]{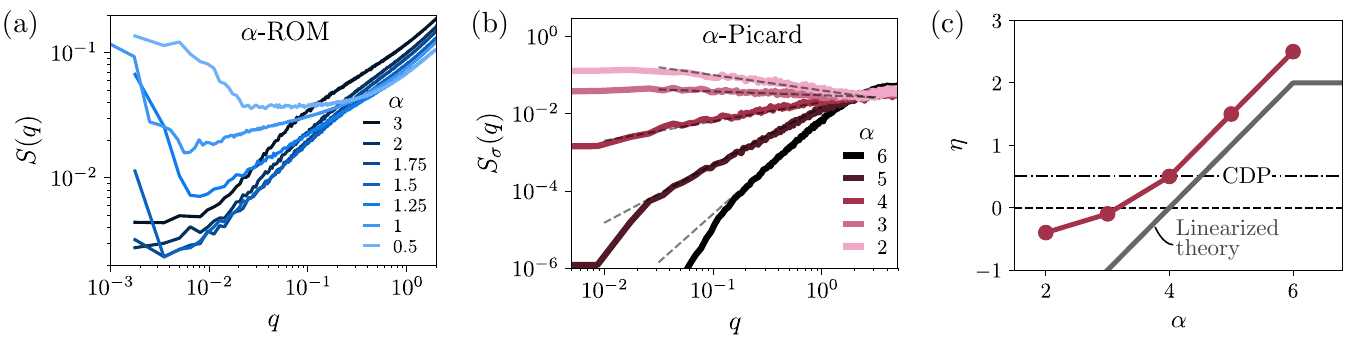}
    \caption{Spatial correlation of the conserved field in Fourier space. (a) Structure factor $S(q)$ in the $\alpha$-ROM. Data originally from~\cite{jocteurRandom2025}. (b) Fourier correlation $S_{\sigma}(q)$ of the stress field in the $\alpha$-Picard model. Fits to power laws $S_\sigma(q)\sim q^\eta$ are shown in dashed lines. (c) Exponents $\eta$ as a function of the power-law decay exponent $\alpha$ of the propagator in the $\alpha$-Picard model. The dashed line at $\eta=0$ bounds hyperuniformity, which is obtained for $\eta>0$. The thick grey line indicates the prediction of the approximate linearized theory (see text).}
    \label{fig:hyperunformity}
\end{figure}

In Fig.~\ref{fig:hyperunformity}(a) we show the measured $S(q)$ in the $\alpha$-ROM for several values of $\alpha$, obtained in steady state close to the critical point, at $\phi-\phi_c(\alpha) \approx 10^{-2}$ for all $\alpha$.
We can observe behaviors compatible with a hyperuniform critical point, with $S(q)$ continuously decreasing when $q$ decreases, for large values of $\alpha \gtrsim 2$, recalling that for $\alpha \geq 3$ the model is in the CDP limit~\cite{jocteurRandom2025}.
However, the presence of longer ranged interactions modifies the picture. For $\alpha < 2$, the structure factor has a non-monotonic behavior as a function of $q$, as $S(q)$ rises for very small $q$ \cite{jocteurRandom2025}.
This signals a loss of hyperuniformity for long-range interactions, again occurring concurrently to the change from a concave to a convex transition.

For elastoplastic models, the stress field is conserved, and the control parameter of the transition is the average stress, meaning that the stress field is the relevant conserved field from the point of view of CDP field theories. 
As a result, a behavior similar to that of the structure factor $S(q)$ in the $\alpha$-ROM may be expected from the spatial correlation of the stress field in the $\alpha$-Picard model. 
By analogy with the structure factor, we consider the Fourier transform of the spatial correlation of the stress field. 
Because the propagator is anisotropic, so is the stress-stress correlation, which depends on the direction of the wavevector $\bm{q}$. 
For simplicity, here we focus on the $q_x=q_y$ direction, and this is the quantity we plot as $S_{\sigma}(q)$ in Fig.~\ref{fig:hyperunformity}(b).
Although some significant differences are visible, the behavior of $S_{\sigma}(q)$ shares some similarities with that of $S(q)$ in the $\alpha$-ROM, with in particular a noticeable $\alpha$-dependence of both $S(q)$ and $S_{\sigma}(q)$ over an intermediate range of wavenumber $q$. A low-$q$ increase of $S_{\sigma}(q)$ is not visible, contrary to $S(q)$. 
However, there is a clear power-law behavior at low $q$, $S_\sigma(q)\sim q^\eta$, with $\eta>0$ at large $\alpha$ values, indicating hyperuniformity, turning to $\eta=0$ at small $\alpha$ values, the change occurring around $\alpha=3$.
Thus, just like for the $\alpha$-ROM, hyperuniformity is lost when the transition is convex.

In the case of yielding, we can measure a hyperuniformity exponent $\eta$ for each $\alpha$, such that for low $q$ values,  $S_\sigma(q)\sim q^\eta$. 
The corresponding power-law behaviors are shown in dashed line in Fig.~\ref{fig:hyperunformity}(b).
Determined values of $\eta$ are reported in Fig.~\ref{fig:hyperunformity}(c), which shows the disappearance of hyperuniformity for $\alpha\lesssim 3$. 
The evolution of the hyperuniformity exponent $\eta$ is rather continuous, and we recover the CDP/short-range depinning value $\eta\approx 0.5$ for $\alpha=4$~\cite{hexnerHyperuniformityCriticalAbsorbing2015}, as well as a value close to the mean-field one $\eta =0$ for $\alpha=3$. 
These two values of $\alpha$ also correspond to the ones separating short-range and mean-field regimes in CDP with long-range transport (LR-CDP)~\cite{lepriolSpatialClusteringDepinning2021}, suggesting that hyperuniformity in the case of yielding evolves exactly according to the LR-CDP scenario for $3<\alpha<4$.
However, in the LR-CDP scenario, the hyperuniformity exponent $\eta$ would saturate to its short-range value $\eta\approx 0.5$ for $\alpha>4$,
while we observe that $\eta$ continuously increases above this value for $4<\alpha<6$.
This strong discrepancy with the LR-CDP scenario for $4<\alpha<6$ is consistent with earlier claims that the presence of zero modes in the propagator shifts the short-range limit up to $\alpha=6$, and induces for $\alpha>6$ a potentially distinct universality class, tentatively called CDP-0~\cite{jocteurYieldingAbsorbingPhase2024a}. While the effect of zero modes on observables directly related to activity seems mild, their effect on hyperuniformity turns out to be strong, which provides further numerical evidence for the existence of a new universality class. This can be understood from the fact that zero modes act on the dynamics of the conserved field (here the local shear stress), thereby directly impacting the spatial correlations of this field.

An approximate analytical determination of $S_{\sigma}(q)$ has been proposed in~\cite{jocteurYieldingAbsorbingPhase2024a} using linearized Langevin equations for the stress and activity fields. The approach was based on a direct coarse-graining of the mesoscopic dynamics in order to derive the evolution equation for the stress field, while using the minimal assumption that the dynamics of the activity field remains the same as in CDP. Within this approximate linearized framework, we obtained $S_{\sigma}(\bm{q}) \sim |\tilde{\mathcal{G}}(\bm{q})|/q^2$ at the critical point, where $\tilde{\mathcal{G}}(\bm{q})$ is the effective macroscopic propagator in Fourier space, satisfying
$\tilde{\mathcal{G}}(\bm{q}) \sim -q_x^2 q_y^2/q^{6-\alpha}$ for $1<\alpha<6$ and $\tilde{\mathcal{G}}(\bm{q}) \sim -q_x^2 q_y^2$ for $\alpha \ge 6$. Restricting ourselves to $q_x=q_y$ as above, we thus get $S_\sigma(q)\sim q^{\alpha-4}$ for $1<\alpha<6$ and $S_\sigma(q)\sim q^2$ for $\alpha \ge 6$. The corresponding prediction for $\eta$ is plotted in Fig.~\ref{fig:hyperunformity}(c), and shows a reasonably good agreement with numerical values of $\eta$ in particular for $4<\alpha<6$. The main interest of this linearized theory prediction is to emphasize the role of the zero modes, which extend the $\alpha$-dependence of $\eta$ up to $\alpha=6$, beyond which a new short-range behavior is found.

\section{Theoretical frameworks}
\label{sec:theoretical:frameworks}

After reviewing in Sec.~\ref{sec:similar:crit} the salient numerical results regarding the similarity of the critical behavior in the $\alpha$-ROM and $\alpha$-Picard models,
we now turn to a brief discussion of tentative theoretical approaches which may allow the critical behavior of both models to be rationalized within common analytical frameworks.
We first briefly discuss the long-range CDP (LR-CDP) field equations, and then turn to a mean-field theoretical framework describing internal noise induced by long-range mediated interactions, and based on a generalization of the H\'ebraud-Lequeux (HL) model for the yielding transition
\cite{hebraudModeCouplingTheoryPasty1998,agoritsasRelevanceDisorderAthermal2015,linMeanFieldDescriptionPlastic2016,linMicroscopicProcessesControlling2018}.

\subsection{Long-Range-CDP field equations}

A first approach consists in generalizing the short-range CDP field equations characterizing the dynamics of the particle density and activity fields to include long-range transport.
In this large-scale continuum description, long-range transport may be mathematically described by adding to the usual Laplacian (valid for short-range transport)
fractional derivative terms \cite{hinrichsenNonequilibriumPhaseTransitions2007}.
These additional transport terms lead to the following evolution equations for the conserved density field $\rho(\bm{r},t)$ and non-conserved activity field $a(\bm{r},t)$ \cite{hinrichsenNonequilibriumPhaseTransitions2007,janssenFieldTheoryDirected2008}:
\begin{align}
\partial_t \rho & = D_\mathrm{SR} \nabla^2 a - D_\mathrm{LR}|\nabla|^{\alpha^\prime-d} a \label{eq:LR_CDP_rho}\\
\partial_t a    & = (r_0+r_1 \rho) a - s a^2  + D_\mathrm{SR} \nabla^2 a - D_\mathrm{LR}|\nabla|^{\alpha^\prime-d} a + \sigma \sqrt{a}\eta\, ,
\label{eq:LR_CDP_a}
\end{align}
where the constant parameters $D_\mathrm{SR}$, $D_\mathrm{LR}$, $r_0$, $r_1$, $s$ and the exponent $\alpha'$ take positive values.
The fractional derivative operator $|\nabla|^{\alpha^\prime-d}$ is defined in Fourier space as
$|\nabla|^{\alpha^\prime-d} \hat{f}(\bm{k}) = |\bm{k}|^{\alpha^\prime-d} \hat{f}(\bm{k})$,
where $\hat{f}(\bm{k})$ is the Fourier transform of an arbitrary field $f(\bm{r})$.
Standard short-range CDP field equations are recovered for $D_{\mathrm{LR}}=0$~\cite{menonUniversalityClassReversibleirreversible2009}. 
On general grounds, the transport coefficients $D_{\mathrm{SR}}$ and $D_{\mathrm{LR}}$ may take different values in the equations for $\rho$ and for $a$.
However, in simple particle models in which activity is carried by particles, transport coefficients for both fields are found to coincide \cite{wieseCoherentstatePathIntegral2016}.

The fractional derivative terms may be interpreted as resulting from long-range random displacements $\bm{u}$ with a power-law distribution
$P(\bm{u}) \sim |\bm{u}|^{-\alpha^\prime}$.
The exponent $\alpha'$ therefore characterizes long-range transport, which is a priori a different process from the long-range induced noise resulting from the power-law propagator with exponent $\alpha$ appearing in particular in the $\alpha$-ROM; hence the use of a distinct notation $\alpha'$ instead of $\alpha$.

\begin{figure}[t]
    \centering
    \includegraphics[width=\textwidth]{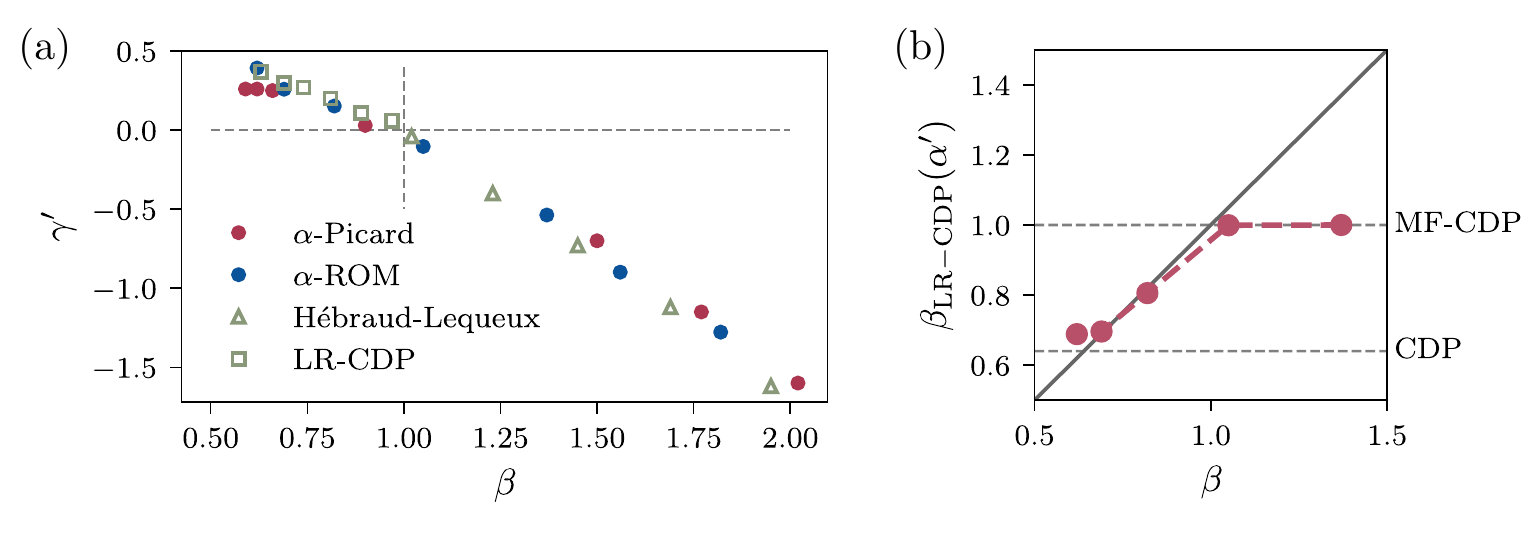}
    \caption{(a) Parametric plot of $\gamma'$ versus $\beta$ in the $\alpha$-ROM and $\alpha$-Picard models as in Fig.~\ref{fig:parametric_exponents}(a), now also including the predictions of the LR-CDP field equations, and of the $\mu$-HL model. The predictions of these two models cover the ranges $\beta_{\text{CDP}}<\beta<1$ and $1<\beta<2$ respectively. The dashed lines indicate the separations between convex and concave transitions at $\beta=1$ and vanishing or diverging critical fluctuations at $\gamma'=0$.
    (b) Effective exponent $\beta_{\mathrm{LR-CDP}}(\alpha')$ of the LR-CDP class, with $\alpha'=d_f+\nu_{\perp}^{-1}$ (scaling relation resulting from statistical tilt symmetry), versus the measured $\beta$ in the $\alpha$-ROM. Horizontal dashed lines indicate the CDP value $\beta\approx 0.64$~\cite{lubeckUniversalScalingBehavior2004} and the mean-field (C)DP value $\beta =1$.
    The solid line is $x=y$.}
    \label{fig:gamma_vs_beta}
\end{figure}

We measured the critical exponents $\beta$ and $\gamma'$ of the class described by the LR-CDP field equations (\ref{eq:LR_CDP_rho}) and (\ref{eq:LR_CDP_a}). Exponent values are plotted in Fig.~\ref{fig:LRCDP:exponents} (see Appendix~\ref{app:LRCDP}) as a function $\alpha'$.
These critical exponents have been measured by using a version of the ROM model with long-range hopping of active particles, which we call LR-ROM. More precisely, in the LR-ROM the probability to jump from position $\bm{r}$ to $\bm{r}+\Delta\bm{r}$ decays as $\sim 1/|\Delta\bm{r}|^{\alpha'}$.
The LR-ROM model is thus a representative of the LR-CDP class.
Note that at variance with the $\alpha$-ROM model, the LR-ROM does include long-range transport of particles.

The fact that the relation, if any, between the exponent $\alpha'$ of the continuum description Eq.~(\ref{eq:LR_CDP_a}) and the exponent $\alpha$ of the microscopic propagator of
the $\alpha$-ROM and $\alpha$-Picard models is a priori unknown makes a direct comparison between the critical properties of the continuum equations and of the microscopic models difficult.
However, it is still possible to plot relations between exponents in a parametric way as in Fig.~\ref{fig:parametric_exponents}.
We reproduce in Fig.~\ref{fig:gamma_vs_beta}(a) data from Fig.~\ref{fig:parametric_exponents}(a), together with the parametric plot $\gamma'$ versus $\beta$ obtained for the LR-ROM, taken as a representative of the LR-CDP class.
Data for the LR-CDP class fall onto the same parametric curve as data from the $\alpha$-ROM and $\alpha$-Picard models.
Yet, as seen in Fig.~\ref{fig:gamma_vs_beta}(a), the LR-CDP exponents only cover the range $\beta<1$ and $\gamma'>0$ (corresponding to a concave transition with diverging fluctuations), while the $\alpha$-ROM and $\alpha$-Picard models cover a broader range of values, including the convex transition characterized by $1<\beta<2$ and $\gamma'<0$. Therefore, to account for this convex transition, an alternative theoretical approach is required, as discussed below.

The fact that the $\alpha$-Picard model is well described over the range $\beta \le 1$ by the LR-CDP class is plausible, since long-range interactions in the $\alpha$-Picard model transport the conserved stress. However, this result is more surprising for the $\alpha$-ROM,
because this model involves only short-range transport, and long-range interactions boil down to mediated noise without long-range transport of the conserved quantity associated with creation of activity.
To further test the idea that the $\alpha$-ROM may be described by an effective long-range transport with an exponent $\alpha' \ne \alpha$,
we use the scaling relation $\nu_{\perp}^{-1}=\alpha-d_f$ valid in the case of the depinning transition with long-range interactions $\sim r^{-\alpha}$ \cite{lin_scaling_2014} (which maps to LR-CDP \cite{le2015exact,wieseTheoryExperimentsDisordered2022}), and derived assuming statistical tilt symmetry. This scaling relation therefore allows us to define an effective exponent $\alpha'=d_f+\nu_{\perp}^{-1}$, from the numerical knowledge of $d_f$ and $\nu_{\perp}$ (the latter being obtained assuming the hyperscaling relation). We then compare in Fig.~\ref{fig:gamma_vs_beta}(b) the exponent $\beta_{\mathrm{LR-CDP}}(\alpha')$ of the LR-CDP class obtained for this effective value of the transport exponent $\alpha'$, to the measured value of $\beta$ in the $\alpha$-ROM. We find a good agreement over the range $\beta \leq 1$, which seems to confirm the relevance of the LR-CDP approach even in the case of the $\alpha$-ROM, provided an effective transport exponent $\alpha'$ can be identified.
However, as mentioned above, the LR-CDP class does not describe the range $1<\beta<2$ (as also visible on Fig.~\ref{fig:gamma_vs_beta}), and we thus now discuss an alternative approach to describe this range of $\beta$ values.

\subsection{Activity creation by mediated noise: generalized H\'ebraud-Lequeux model}

\subsubsection{Mediated noise statistics}
A key mechanism responsible for the onset of a convex transition is the presence of internal mediated noise induced by distant active events through long-range mediated interactions.
This mechanism has been known for long in the frame of the elastoplastic scenario for the yielding transition
\cite{hebraudModeCouplingTheoryPasty1998,picardElasticConsequencesSingle2004,nicolas_deformation_2018}, and has been recently argued to play a similar role for the $\alpha$-ROM model
for long-enough interaction range \cite{jocteurRandom2025}.
In this section, we briefly discuss this mediated noise scenario, mostly based on generalizations of the H\'ebraud-Lequeux model for the yielding transition
\cite{hebraudModeCouplingTheoryPasty1998,agoritsasRelevanceDisorderAthermal2015,linMeanFieldDescriptionPlastic2016,linMicroscopicProcessesControlling2018},
which is a simple mean-field model for the diffusion over a threshold induced by mediated noise.

The precise meaning of mediated noise may slightly depend on the model considered. For instance, in the $\alpha$-ROM, the mediated noise corresponds to random displacements
of passive particles induced by the distant motion of active particles \cite{jocteurRandom2025}.
In the $\alpha$-Picard model, the mediated noise rather refers to the (possibly anomalous~\cite{linDensityShearTransformations2014}) diffusion of the local stress, induced by distant plastic events \cite{nicolas_deformation_2018,jocteurYieldingAbsorbingPhase2024a}.
We generically denote as $\delta\xi$ the mediated noise induced on a given location by distant events. It may correspond to the noise acting on the local stress in the $\alpha$-Picard model, or to one of the components of the displacement vector of passive particles in the $\alpha$-ROM model.
In both cases, the distribution $p(\delta\xi)$ of the mediated noise $\delta\xi$ shares similar generic properties, which depend on the decay exponent $\alpha$ of long-range mediated interactions. An analysis of the dependence of the noise distribution $p(\delta\xi)$ on $\alpha$ may be performed at a mean-field level \cite{jocteurRandom2025}.
For $\alpha<d/2$, $p(\delta\xi)$ is a Gaussian distribution with a variance proportional to the mean activity $A$ in the system \cite{jocteurRandom2025}.
By contrast, when $\alpha>d/2$, the noise distribution $p(\delta\xi)$ takes a power-law tail for large $|\delta\xi|$
\cite{linDensityShearTransformations2014,linMeanFieldDescriptionPlastic2016,linMicroscopicProcessesControlling2018,ferreroCriticalityElastoplasticModels2019,jocteurRandom2025},
\begin{equation}
    p(\delta\xi) \sim \frac{A}{|\delta\xi|^{1+\mu}},
    \label{eq:noise:dist}
\end{equation}
where the prefactor $A$ is again the mean activity.
The exponent $\mu$, which satisfies $0<\mu<2$, defines the broadness of the noise, and plays a key role in the critical properties of the absorbing phase transition, as discussed below.
In the present mean-field analysis, one finds $\mu=d/\alpha$, with $d$ the space dimension \cite{jocteurRandom2025}.

Beyond the mean-field level of analysis, one may still expect the mediated noise distribution to take a Gaussian form for small values of $\alpha$, and a power-law form
similar to Eq.~(\ref{eq:noise:dist}) for larger $\alpha$. However, the dependence of the exponent $\mu$ on $\alpha$ may differ from the simple mean-field form $\mu=d/\alpha$,
although we presently have no analytical prediction for $\mu$ beyond the mean-field one.

\subsubsection{Hébraud-Lequeux model with Gaussian noise}

We may now try to use the above noise statistics in a dynamical model of activity creation, based on the similarities between the activity creation mechanisms appearing
in the $\alpha$-ROM and $\alpha$-Picard models, by generalizing the H\'ebraud-Lequeux model originally introduced in the frame of the yielding transition
\cite{hebraudModeCouplingTheoryPasty1998,agoritsasRelevanceDisorderAthermal2015,bouchaud_spontaneous_2016,linMeanFieldDescriptionPlastic2016,linMicroscopicProcessesControlling2018}.
On general grounds, the basic ingredients of the H\'ebraud-Lequeux model are the following.
Activity is created locally when a local variable $\sigma_i$ overcomes a threshold $\sigma_c$, in which case $\sigma_i$ relaxes.
In elastoplastic models, $\sigma_i$ is the local shear stress. 
The evolution of $\sigma_i$ results from a slow external drive which tends to steadily increase the value of $\sigma_i$, like an external shear in the yielding case, and from a (possibly anomalous) diffusion resulting from the mediated noise, generated by distant creation of activity (i.e., other variables $\sigma_j$ overcoming the threshold $\sigma_c$).
H\'ebraud-Lequeux-type models treat the problem at mean-field level by focusing on a single local variable $\sigma$, self-consistently determining the average activity created remotely from the statistics of $\sigma$ itself.

For a Gaussian noise distribution $p(\delta\xi)$ (corresponding to $\alpha<d/2$ in the mean-field approximation), 
the evolution equation for the probability distribution $P(\sigma,t)$ takes the form \cite{hebraudModeCouplingTheoryPasty1998,agoritsasRelevanceDisorderAthermal2015}:
\begin{equation}
    \frac{\partial P}{\partial t}(\sigma,t) = - g_0 \dot{\gamma} \frac{\partial P}{\partial \sigma}(\sigma,t) + \mathcal{D} \frac{\partial^2 P}{\partial \sigma^2}(\sigma,t) - \frac{1}{\tau}\Theta(|\sigma|-\sigma_c)P(\sigma,t) + A \delta(\sigma)\, ,
    \label{eq:HL:diff}
\end{equation}
where $g_0$ is the shear elastic modulus, and $\Theta(x)$ is the Heaviside function. The quantity $A$ in Eq.~(\ref{eq:HL:diff}) is the average activity defined as
\begin{equation}
    A(t) = \frac{1}{\tau} \int_{|\sigma|>\sigma_c}\mathrm{d}\sigma P(\sigma,t)\, .\label{eq:HL:activity:def}
\end{equation}
Here $\tau$ is the typical time to relax the local stress once $|\sigma|>\sigma_c$.
By a self-consistent mean-field argument, the diffusion coefficient $\mathcal{D}$ in $\sigma$-space is proportional to the average activity 
\begin{equation}
\mathcal{D} = \kappa A\, , \label{eq:HL:self-consist:diff}
\end{equation}
where the proportionality coefficient $\kappa>0$ may be determined from the detailed knowledge of the noise distribution.

While it was originally formulated in the framework of elastoplastic models like the Picard model, the H\'ebraud-Lequeux model can be adapted to the suspension context
under some simplifying assumptions. One may for instance assume that neighboring particles of a given passive particle may be replaced by a circular cage of radius $R$ \cite{jocteurRandom2025}.
In this simplified picture, the focus passive particle diffuses under the effect of the mediated noise until it reaches the cage and becomes active.
Once active, the particle randomly jumps, becomes passive again, and is then assumed to be placed at the center of a new cage.
Note that although the above assumptions may be considered as rough approximations, which may contrast with the elastoplastic case where the modelling assumptions are more straightforward,
it should be kept in mind that elastoplastic models are themselves rough mesoscopic approximations of more realistic underlying particle systems.
In the suspension context, the local variable to be considered (instead of $\sigma$) is the displacement vector $\bm{r}$ from the center of the cage.
The particle becomes active when $||\bm{r}|| \equiv r > R$.
For a Gaussian mediated noise, inducing a normal diffusion of the passive particle with diffusion coefficient $\mathcal{D}$, the distribution $P(\bm{r},t)$ evolves according to
\cite{jocteurRandom2025}
\begin{equation}
    \partial_t P(\bm{r},t) = \mathcal{D}\nabla^2 P(\bm{r},t) - \frac{1}{\tau}\Theta(r-R)P(\bm{r},t) + A \delta(\bm{r})\, ,
    \label{eq:SHL:diff}
\end{equation}
where the mean activity $A$ is defined as
\begin{equation}
    A(t) = \frac{1}{\tau} \int_{|\bm{r}|>R}\mathrm{d}\bm{r} P(\bm{r},t)\, ,\label{eq:SHL:activity:def}
\end{equation}
$\tau$ being the typical time for a particle to jump when it becomes active.
As in the elastoplastic case, a self-consistent mean-field argument implies that the diffusion coefficient $\mathcal{D}$ is proportional to the average activity
as in Eq.~(\ref{eq:HL:self-consist:diff}).
Beyond the vectorial nature of the variable $\bm{r}$, the main difference with the H\'ebraud-Lequeux model defined in an elastoplastic context is the absence of a drift term in Eq.~(\ref{eq:SHL:diff}), which was linked to the shear rate $\dot{\gamma}$ in the elastoplastic version of the model described by Eq.~(\ref{eq:HL:diff}).
Note that although the model is primarily considered in two dimensions, the vectorial form of Eq.~(\ref{eq:SHL:diff}) is valid in arbitrary space dimension.

In both the yielding and suspension cases, one finds an absorbing phase transition between an active phase with $A>0$ and an absorbing one with $A=0$, with critical exponent $\beta=2$.
However, the control parameters differ in both versions of the H\'ebraud-Lequeux model. 
For the elastoplastic version, the control parameter is the average stress $\langle \sigma \rangle$, and a transition occurs at a yield value $\Sigma_Y$, such that for $\langle \sigma \rangle>\Sigma_Y$, one has $A \sim (\langle \sigma \rangle -\Sigma_Y)^2$.
In the suspension version, the control parameter is the packing fraction $\phi$ of particles, related to the cage radius $R$ by $\phi \sim R^{-d}$.
One then finds a critical cage radius $R_c$ such that for $R<R_c$, $A \sim (R_c-R)^2 \sim (\phi -\phi_c)^2$ \cite{jocteurRandom2025}.
The physical reason for the emergence of the non-standard value $\beta=2$ of the critical exponent is the diffusion of passive particles above a threshold to create activity,
with a diffusion coefficient self-consistently induced by the presence of activity.
This is most easily illustrated in the elastoplastic case. Over the characteristic time $\tau$, the mediated noise induces a stress diffusion over a stress scale
$\delta \sigma \sim (\mathcal{D}\tau)^{1/2} \propto (A\tau)^{1/2}$, hence the scaling $A \sim (\delta\sigma)^2$.

\subsubsection{H\'ebraud-Lequeux model with heavy-tailed noise}

We now turn to the study of the H\'ebraud-Lequeux model in the opposite situation when the noise is power-law distributed (corresponding to $\alpha>d/2$ in the mean-field scenario),
which leads to a modified form of the evolution equation for the probability distribution of the local variable.
In the elastoplastic case, the evolution equation for $P(\sigma,t)$ now reads \cite{linMeanFieldDescriptionPlastic2016,linMicroscopicProcessesControlling2018}
\begin{equation}
    \frac{\partial P}{\partial t}(\sigma,t) = - g_0 \dot{\gamma} \frac{\partial P}{\partial \sigma}(\sigma,t) - \mathcal{D} |\partial_{\sigma} |^{\mu} P(\sigma,t) - \frac{1}{\tau}\Theta(|\sigma|-\sigma_c)P(\sigma,t) + A \delta(\sigma)\, .
    \label{eq:HL:anomalous}
\end{equation}
where the anomalous diffusion coefficient $\mathcal{D}$ is still proportional to the mean activity $A$ as in Eq.~(\ref{eq:HL:self-consist:diff}).
This comes from the fact that in the power-law case, the tails of the noise distribution $p(\xi)$ are proportional to the mean activity $A$, as seen on Eq.~(\ref{eq:noise:dist}).
The fractional derivative operator $|\partial_{\sigma}|^{\mu}$ is defined in Fourier space by $|\partial_{\sigma}|^{\mu}\hat{f}(k) = |k|^{\mu}\hat{f}(k)$.
The main difference between Eq.~(\ref{eq:HL:anomalous}) and the Gaussian noise version Eq.~(\ref{eq:HL:diff}) is that the Laplacian is replaced by a fractional derivative.

For the suspension model in the presence of heavy-tailed noise, Eq.~(\ref{eq:SHL:diff}) is modified into \cite{jocteurRandom2025}
\begin{equation}
    \partial_t P(\bm{r},t) = - \mathcal{D}|\nabla|^\mu P(\bm{r},t) - \frac{1}{\tau}\Theta(r-R)P(\bm{r},t) + A \delta(\bm{r})\, ,
    \label{eq:SHL:anomalous}
\end{equation}
with the fractional derivative $|\nabla|^{\mu}\hat{f}(\bm{k}) = |\bm{k}|^{\mu}\hat{f}(\bm{k})$.
For the same reasons as above, the diffusion coefficient follows the self-consistent relation $\mathcal{D}=\kappa A$.

\begin{figure}
    \centering
    \includegraphics[width=\textwidth]{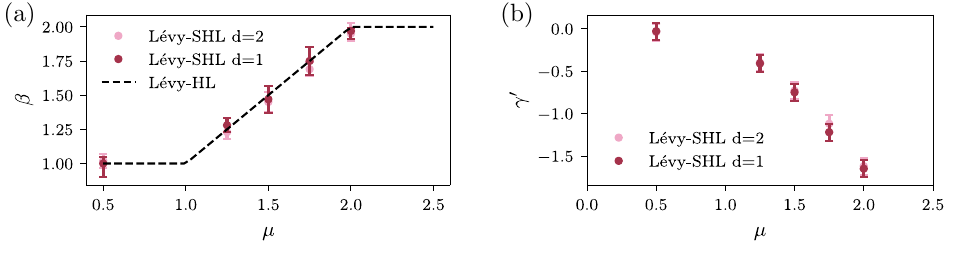}
    \caption{Evolution with the noise exponent $\mu$ of (a) the order parameter exponent $\beta$ (data originally from~\cite{jocteurRandom2025}) and (b) the fluctuations exponent $\gamma^\prime$, in the Lévy-Hébraud-Lequeux model. Dots: numerical resolution of the suspension Lévy-Hébraud-Lequeux model given by Eq.~(\ref{eq:SHL:anomalous}) in space dimension $d=1$ and $d=2$. Dashed line: analytical solution of the Lévy-Hébraud-Lequeux model defined by Eq.~(\ref{eq:HL:anomalous}) \cite{linMeanFieldDescriptionPlastic2016}.}
    \label{fig:HL:beta:gamma:mu}
\end{figure}

Both versions of this generalized H\'ebraud-Lequeux model with anomalous noise lead to a critical exponent $\beta=\mu$ for $1<\mu<2$ and $\beta=1$ for $\mu \le 1$
\cite{linMicroscopicProcessesControlling2018,jocteurRandom2025}.
The exponent $\beta$ measured in the suspension H\'ebraud-Lequeux model in both $d=2$ and $d=1$ is plotted versus the noise exponent $\mu$ in Fig.~\ref{fig:HL:beta:gamma:mu}(a).
We recall that due to the different control parameters in both versions of the model, the critical exponent $\beta$ is defined as $A \sim (\langle \sigma \rangle -\Sigma_Y)^{\beta}$
in the elastoplastic case, and as $A \sim (R_c-R)^{\beta} \sim (\phi-\phi_c)^\beta$ in the suspension case.
In the present case of an anomalous diffusion, the dependence of $\beta$ on $\mu$ can also be understood through a simple heuristic argument, at least for $1<\mu<2$.
Specializing again to the elastoplastic case, the noise-induced anomalous diffusion leads over the characteristic time $\tau$ to a typical stress variation
$\delta \sigma \sim (\mathcal{D}\tau)^{1/\mu} \propto (A\tau)^{1/\mu}$, yielding the scaling $A \sim (\delta\sigma)^{\mu}$ (recalling the assumption $1<\mu<2$), where $\delta \sigma >0$.

To complete our study of the generalized H\'ebraud-Lequeux model, we also determine numerically the exponent $\gamma'$ characterizing fluctuations of activity, $\langle \delta A\rangle \sim \varepsilon^{-\gamma'}$, which requires to define a finite-size version of the H\'ebraud-Lequeux model.
Indeed, the above formulation of several variants of the H\'ebraud-Lequeux model in terms of an evolution equation for the probability distribution of the local variable (i.e., $\sigma$ or $\bm{r}$) is well suited to determine the critical exponent $\beta$, as the latter is defined from the average activity $A$, but not to characterize fluctuations since the model is implicitly defined in the limit of an infinite-size system. To access fluctuations, one thus needs to define a finite-size version of the H\'ebraud-Lequeux model.
In practice, this amounts to considering a large but finite $\alpha$-ROM with a spatially uniform propagator, corresponding to the limit $\alpha \to 0$.
We therefore measured the exponent $\gamma'$ characterizing activity fluctuations by simulating an $\alpha$-ROM composed of $\Ntot =10^7$ particles with a spatially uniform propagator $(\alpha=0)$,
in dimensions $d=2$ and $d=1$. Details of the numerical method can be found in Appendix~D of \cite{jocteurRandom2025}.
The resulting values of $\gamma'$ are plotted as a function of the noise exponent $\mu$ in Fig.~\ref{fig:HL:beta:gamma:mu}(b).
No dependence on the space dimension is observed, as values of $\gamma'$ for $d=2$ and $d=1$ coincide within error bars.
Note that the values of $\beta$ plotted in Fig.~\ref{fig:HL:beta:gamma:mu}(a) were also obtained using the same numerical simulation method, which is technically more convenient than a direct numerical integration of Eq.~(\ref{eq:HL:anomalous}) or (\ref{eq:SHL:anomalous}).

Finally, we draw on Fig.~\ref{fig:gamma_vs_beta}(a) the parametric plot $\gamma'$ versus $\beta$ for the generalized H\'ebraud-Lequeux model, using data from Fig.~\ref{fig:HL:beta:gamma:mu}.
We observe that data from the H\'ebraud-Lequeux model nicely match data from the $\alpha$-ROM or $\alpha$-Picard model in the range $\beta >1$ (convex transition).
Altogether, we see on Fig.~\ref{fig:gamma_vs_beta}(a) that the concave regime $\beta <1$ (with diverging fluctuations, $\gamma'>0$) is well described by the LR-CDP framework, whereas the convex case $\beta>1$ (with vanishing fluctuations, $\gamma'<0$) is well accounted for by the generalized H\'ebraud-Lequeux model approach. It is interesting to observe that both models have no overlap in terms of values of the critical exponents $\beta$ and $\gamma'$.

\section{Discussion}

We have shown that two models describing driven athermal soft matter systems with long-range mediated interactions display unconventional absorbing phase transitions with marked differences in their critical properties with respect to the CDP class which describes short-range interacting systems.
Both long-range interacting models we considered display a convex absorbing phase transition (i.e., with $\beta>1$, where $\beta$ is the order parameter critical exponent),
vanishing critical fluctuations of the order parameter, and normal spatial fluctuations of the conserved field at the critical point (as opposed to hyperuniformity found in short-range systems belonging to the CDP class).
This convex character of the transitions reported in the two models escapes the usual classification of absorbing phase transitions with long-range interactions in terms of the Long-Range Conserved Directed Percolation (LR-CDP) class, therefore raising important conceptual challenges for the general classification of absorbing phase transitions.

From a dynamical standpoint, the intermittent dynamics close to the critical point takes the form of avalanches, which in both long-range models display a non-compact spatiotemporal structure, with a fractal dimension $d_f<d$, again at variance with the short-range case for which $d_f>d$.
Continuously varying the decay exponent $\alpha$ of long-range interactions in both models, we have observed that the crossover from concave ($\beta<1$) to convex ($\beta>1$) transitions coincides,
around a model-dependent value $\alpha^\ast$, with the crossover from diverging to vanishing order parameter fluctuations, and from compact to non-compact avalanches.
This coincidence suggests that geometrical properties of avalanches might also indirectly rule some of the static critical properties of the absorbing phase transition,
as also suggested by the scaling relation $\beta = \nu_\perp (d-d_f+z)$ \cite{lin_scaling_2014} (numerically confirmed in the $\alpha$-ROM \cite{jocteurAvalanches2026})
which relates the static exponents $\beta$ and $\nu_\perp$ to the dynamic exponents $d_f$ and $z$ characterizing avalanches.

Although the overall phenomenology of both models studied here is similar, a significant difference lies in the values of the decay exponent $\alpha$  over which the critical exponents are floating (i.e., continuously depend on $\alpha$).
In particular, the values $\alpha^\ast$ at which the transition qualitatively changes nature from convex to concave is at  $\alpha^\ast \approx 1.5$ for the $\alpha$-ROM and $\alpha^\ast \approx 3$ for the $\alpha$-Picard.
These differences may be interpreted as resulting from different types of long-range interactions: interactions proceed via long-range transport of a conserved quantity (the shear stress) in the $\alpha$-Picard model, whereas transport remains short-ranged but with long-range mediated noise in the $\alpha$-ROM.
Let us outline further that the relevant conserved quantity from the viewpoint of the absorbing phase transition is the one coupled to the creation of activity. Other conserved quantities, not directly coupled to activity, may be present and subject to long-range transport, like stress in suspensions; such conserved quantities have not been explicitly taken into account in the models studied.

Quite remarkably, however, these apparent quantitative differences seem to vanish when the floating exponents are plotted parametrically, which reveals that the $\alpha$-ROM and $\alpha$-Picard models fall on a shared master curve for $\beta > 0.64$ (i.e., when zero modes play no role \cite{jocteurYieldingAbsorbingPhase2024a}).
Whether the a priori different nature of long-range interactions (and thus different ranges of relevant $\alpha$ values) for the two models actually veils that they belong to the same class is an enthralling speculation, but remains to be backed by a theoretical scenario beyond our numerical observations. 
A possibility is that there exists a mapping relating the natural observables of both models, much like there is a mapping between disordered elastic interfaces and fixed-energy sandpiles~\cite{le2015exact}.

An extra piece to this discussion is our observation that the master curve revealed by the parametric plot of exponents is entirely covered, in two complementary pieces, by LR-CDP, which is CDP with long-range transport of the conserved quantity, and generalized Hébraud-Lequeux models, which make a mean-field description for long-range mechanical noise on the conserved quantity.
LR-CDP visits the entire part of the master curve associated with concave transitions when varying the exponent $\alpha'$ associated with long-range transport, 
while generalized HL visits the entire part associated with convex transition when varying the noise exponent $\mu$, which corresponds to varying the exponent of long-range mechanical noise $\alpha$ as suggested by the mean-field estimate $\mu=d/\alpha$.
This points to the possibility of a unified description covering both APTs governed by long-range transport and APTs governed by long-range mechanical noise, an idea already put forward in the context of yielding~\cite{lin_scaling_2014,ferreroCriticalityElastoplasticModels2019,ferreroElasticInterfacesDisordered2019}.

If it exists, such unified description would come with several interesting features and raise questions.
First, anomalous diffusion or transport in the LR-CDP and generalized HL frameworks play different roles.
In the LR-CDP case, real-space transport of a conserved quantity occurs due to a long-range propagator $\mathcal{G}(r) \sim 1/r^{\alpha'}$.
For $\alpha'>d+2$, transport is diffusive and the dynamics effectively remains short-ranged, while for $\alpha'<d+2$ anomalous transport sets in, thereby entering the true LR-CDP regime.
In the HL case, diffusion is induced by mediated noise resulting on a given location from all distant active events, mediated by the long-range propagator. 
Diffusion occurs here in a somewhat different space, either the locally conserved variable like the local stress in the $\alpha$-Picard model, or in real-space but on very short length scales like in the $\alpha$-ROM (as opposed to the large length scales on which transport occurs in the LR-ROM framework).
Diffusion in this `auxiliary space' due to mediated noise becomes important because it takes place close to a threshold for creation of activity.
The criterion for anomalous diffusion due to mediated noise is different from the one for anomalous transport in the LR-ROM. 
Anomalous diffusion takes place when $\mu<2$, where $\mu$ is the power-law exponent of the noise distribution $p(\delta\xi)$ [see Eq.~(\ref{eq:noise:dist})], which results in the condition $\alpha>d/2$ under the mean-field estimate $\mu=d/\alpha$.
Interestingly, anomalous diffusion occurs in the HL case for large enough $\alpha$, meaning for sufficiently short ranged noise propagation, while it occurs for small enough $\alpha'$ in LR-CDP, that is, for sufficiently long ranged transport.

Second, there is a tension around the mean-field nature (as opposed to finite-dimensional) of the transition. 
For LR-CDP, a mean-field behavior is recovered at very long range, when $\alpha'<3d/2$.
In two dimensions, this occurs at $\alpha'=3$, which is in good agreement with our data for the two-dimensional $\alpha$-Picard model, showing a mean-field DP-like behavior for $\alpha=3$, with $\beta\approx 1$, $\gamma' \approx 0$, $\tau \approx 1.5$, $d_f\approx 2$, $\chi\approx 2$, $z\approx 1$ (note however that the critical decay exponent $\delta$, $A(t)\sim t^{-\delta}$ takes a value $\delta \approx 0.5$ instead of $\delta_{MF}=1$ as mentioned above).
For the $\alpha$-ROM model, as already discussed it is not possible to readily read the $\alpha$ as a transport exponent, nonetheless there is a point at $\alpha=1.5$ which shares the same mean-field DP features.
Thus,  for both model, everything seems to be compatible with reaching the mean-field DP limit when gradually increasing the range of interactions from the short-range regime.
However, this mean-field-like point is destabilized by a further increase of the interaction range, entering a HL regime where the exponent evolution is controlled by long-range mechanical noise rather than long-range transport.
Yet, this regime is entirely mean-field only when the mechanical noise exponent $\mu$ is large enough, $\mu\geq 2$, i.e. when it gives rise to regular (not anomalous) diffusion.
This happens for the longest ranges of mechanical noise. 
In the anomalous diffusion regime, the mean-field nature partially disappears, in the sense that only a part, but not all of the scaling relations follow their mean-field form.
This has been observed for both the $\alpha$-ROM~\cite{jocteurRandom2025} and the Picard model~\cite{ferreroElasticInterfacesDisordered2019}.

Third, the mathematical representations of LR-CDP and generalized HL models fundamentally differ.
LR-CDP naturally lends itself to a field theoretic description involving two coupled scalar fields, namely an non-conserved activity field and a conserved density field~\cite{rossi2000universality,menonUniversalityClassReversibleirreversible2009}.
By contrast, the HL model considers the dynamics of the distribution of the conserved field, without allowing a reduction to a scalar dynamics that would retain the phenomenology.
This fundamentally is related to the fact that in the HL model the physics near the transition is controlled by the diffusion in the auxiliary space, near the threshold to create activity, which cannot be captured by a moment of the distribution.

A unified picture should capture this intricate phenomenology. 
A key question would be to confirm and to understand from a coarse-grained field theory the seeming emergence of effective long-range transport in the $\alpha$-ROM.
This tentative scenario is suggested by the scaling relation $\nu_{\perp}^{-1}=\alpha-d_f$ \cite{lin_scaling_2014}, which although not satisfied in the $\alpha$-ROM, allows for the definition of an effective value $\alpha_{\mathrm{eff}}$. Once plugged into the long-range CDP class, $\alpha_{\mathrm{eff}}$ predicts the correct value of the exponent $\beta$ (i.e., the one measured in the $\alpha$-ROM) as long as $\beta\le 1$. Although a loophole cannot be excluded to escape the above argument, a better understanding of the potential correspondence between long-range transport and long-range mediated noise without transport would be useful.
In addition, the aforementioned scaling relation $\nu_{\perp}^{-1}=\alpha-d_f$ is known to result from a statistical tilt symmetry in the depinning context.
Whether an equivalent symmetry to the statistical tilt symmetry is present in the $\alpha$-ROM is also a question of interest, which might help deepening the connection between the depinning and reversible-irreversible transitions.
It is unclear to us whether a putative unified picture is possible using a generalization of the continuum description for LR-CDP. 
Possibly such generalization would involve a distribution field (in the spirit, e.g., of the KEP model \cite{bocquetKinetic2009}), not only scalar fields.

\bigskip
\ack{The authors are grateful to Lasse Laurson, Pierre Le Doussal, Cesare Nardini, and St\'ephane Santucci for stimulating discussions.}

\funding{This project was provided with computer and storage resources by GENCI at
IDRIS thanks to the grants 2023-AD010914551, 2024-AD010914551R1 and 2025-AD010914551R2 on the supercomputer Jean Zay's V100 and A100 partitions. 
Some of the computations presented in this paper were performed using the GRICAD infrastructure (\href{https://gricad.univ-grenoble-alpes.fr}{https://gricad.univ-grenoble-alpes.fr}), which is supported by Grenoble research communities. T.J. acknowledges funding from the French Ministry of Higher Education and Research.}

\data{Data sets generated during the current study are available from the corresponding author upon reasonable request.}

\bigskip

\appendix

\section{Numerical determination of the critical exponents of the LR-ROM}
\label{app:LRCDP}

\begin{figure}[h]
    \centering
    \includegraphics[width=0.6\textwidth]{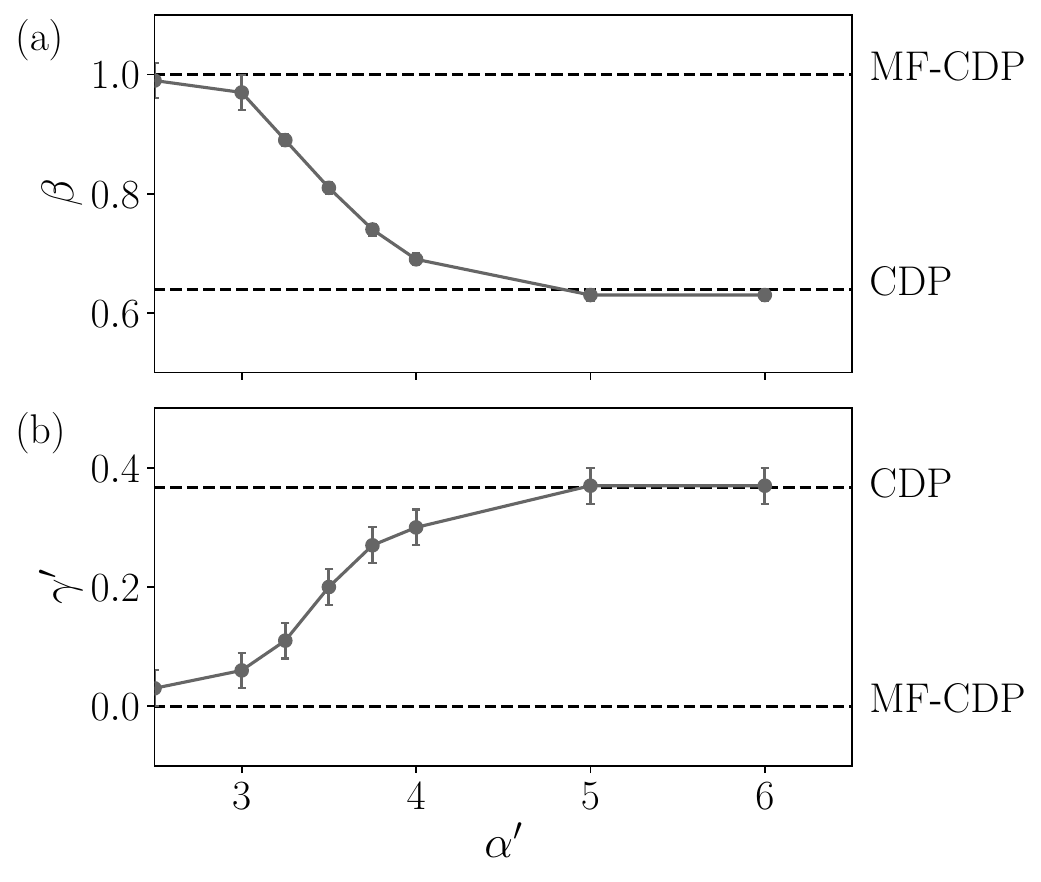}
    \caption{Long-range conserved directed percolation (LR-CDP) exponents as function of the power-law transport exponent $\alpha'$. (a) Order parameter exponent $\beta$. (b) Order parameter fluctuations exponent $\gamma'$.}
    \label{fig:LRCDP:exponents}
\end{figure}

We have determined numerically the critical behavior of the LR-ROM model, which falls into the LR-CDP class.
We focus specifically on the three exponents $\beta$, and $\gamma'$, respectively defined by the scaling relations
$\langle A\rangle \sim \delta\phi^{\beta}$ and $\langle (\delta A)^2 \rangle \sim \delta\phi^{-\gamma'}$.
These two exponents are plotted in Fig.~\ref{fig:LRCDP:exponents} as a function of the exponent $\alpha'$ characterizing the long-range transport, as defined in Eqs.~(\ref{eq:LR_CDP_rho},\ref{eq:LR_CDP_a}).

\bibliographystyle{unsrt.bst}
\bibliography{biblio.bib}

\end{document}